\documentclass[rnote]{aa}
\usepackage{txfonts}
\usepackage{graphicx}
\usepackage{latexsym}

\begin{document}

\title{
Determination of the cosmic far-infrared background level with the ISOPHOT
instrument
\thanks{
Based on observations with the Infrared Space Observatory ISO. ISO is an ESA
project with instruments funded by ESA member states (especially the PI
countries France, Germany, The Netherlands, and the UK) and with participation
of ISAS and NASA.
}}

\author{M. Juvela  \inst{1}
\and   K. Mattila\inst{1}
\and D. Lemke\inst{2}
\and U. Klaas\inst{2}
\and C. Leinert\inst{2}
\and Cs. Kiss\inst{3}
}

\offprints{M. Juvela}

\institute{Observatory, University of Helsinki,
P.O. Box 14, FI-00014 Helsinki, Finland
\and
Max-Planck-Institut f\"ur Astronomie,
K\"onigstuhl 17, D-69117 Heidelberg, Germany
\and
Konkoly Observatory of the Hungarian Academy of Sciences, 
P.O. Box 67, H-1525 Budapest, Hungary
}

\date{Received 1 January 2005 / Accepted 2 January 2005}

\abstract
{The cosmic infrared background (CIRB) consists mainly of the integrated light
of distant galaxies. In the far-infrared the current estimates of its surface
brightness are based on the measurements of the COBE satellite. Independent
confirmation of these results is still needed from other instruments.}
{In this paper we derive estimates of the far-infrared CIRB using 
measurements made with the ISOPHOT instrument aboard the ISO satellite. The
results are used to seek further confirmation of the CIRB levels that have
been derived by various groups using the COBE data. }
{We study three regions of very low cirrus emission. The surface brightness
observed with the ISOPHOT instrument at 90, 150, and 180$\mu$m is correlated
with hydrogen 21\,cm line data from the Effelsberg radio telescope.
Extrapolation to zero hydrogen column density gives an estimate for the sum of
extragalactic signal plus zodiacal light. The zodiacal light is subtracted
using ISOPHOT data at shorter wavelengths. Thus, the resulting estimate of the
far-infrared CIRB is based on ISO measurements alone. }
{In the range 150 to 180$\mu$m, we obtain a CIRB value of
1.08$\pm$0.32$\pm$0.30\,MJy\,sr$^{-1}$ quoting statistical and systematic
errors separately. In the 90\,$\mu$m band, we obtain a 2-$\sigma$ upper limit
of 2.3\,MJy\,sr$^{-1}$. }
{ The estimates derived from ISOPHOT far-infrared maps are consistent with the
earlier COBE results. 
}


\maketitle

\section{Introduction}

The extragalactic background light (EBL) consists of the integrated light of
all galaxies along the line of sight with possible additional contributions
from intergalactic gas and dust and hypothetical decaying relic particles. It
plays an important role in cosmological studies because most of the
gravitational and fusion energy released in the universe since the
recombination epoch is expected to reside in the EBL. Measurements of the
cosmic infrared background, CIRB, help to address some central, but still
largely open astrophysical problems, including the early evolution of
galaxies, and the entire star formation history of the universe. An important
issue is the balance between the UV-optical-NIR and the far-infrared
backgrounds; the fraction of optical radiation lost by dust obscuration
re-appears as dust emission at longer wavelengths. The absolute level of
the CIRB, the fluctuations in the CIRB surface brightness, and the resolved
bright end of the distribution of galaxies contributing to the CIRB all
provide strong constraints on the models of galaxy evolution through different
epochs. For reviews, see Hauser \& Dwek (\cite{Hauser2001}) and Lagache,
Puget, \& Dole (\cite{Lagache2005}).
The full analysis of the data from the DIRBE (Hauser et al. \cite{Hauser1998};
Schlegel et al. \cite{schlegel}) and FIRAS (Fixen et al. \cite{fixen})
experiments indicated a CIRB at a surprisingly high level of $\sim$1
MJy\,sr$^{-1}$ between 140 and 240 $\mu$m. Preliminary results had been
obtained by Puget et al. (\cite{puget}). Lagache et al. (\cite{Lagache1999})
claimed the detection of a component of Galactic dust emission associated with
warm ionised medium. The removal of this component led to a CIRB level of 0.7
MJy\,sr$^{-1}$ at 140 $\mu$m.

Because the FIR CIRB is important for cosmology these results need to be
confirmed by independent measurements.  The ISOPHOT instrument (Lemke at el.
\cite{Lemke96}), flown on the cryogenic, actively cooled ISO satellite,
provided the capabilities for this. The ISOPHOT observation technique was
different from COBE: (1) with its relatively small f.o.v. ISOPHOT was capable
of looking into the darkest spots between the cirrus clouds; (2) ISOPHOT had
high sensitivity in the important FIR window at 120-200 $\mu$m; (3) with its
good spatial and multi-wavelength FIR spectral sampling ISOPHOT gave an
improved possibility of separating and eliminating the emission of Galactic
cirrus. 
The primary goal of the ISOPHOT EBL project
is the determination of the absolute level of the FIR CIRB. The other goals
are the measurement of the spatial CIRB fluctuations and the detection of the
bright end of the FIR point source distribution. The bright end of the galaxy
population contributing to the FIR CIRB signal was analysed by Juvela et al.
(\cite{Juvela00}).

\section{The method}

We examine three regions of low cirrus emission that were mapped with the
ISOPHOT 
at 90, 150, and 180$\mu$m. 
Because of the high sensitivity of the ISOPHOT FIR
detectors, we can directly correlate HI with ISOPHOT measurements for each FIR
band separately. In the case of DIRBE, 
the original analysis performed by the DIRBE team used 100\,$\mu$m as an ISM template
and, therefore, the accuracy of the CIRB detections at 140\,$\mu$m and
240\,$\mu$m also depended on the systematic uncertainties of the 100\,$\mu$m
data (Hauser et al.~\cite{Hauser1998}; Arendt et al. \cite{Arendt1998}).

The HI lines are optically thin and their intensity traces the amount of
neutral hydrogen along the line-of-sight. 
The level of FIR emission associated with the ionised
medium is still uncertain and we will consider the possible effects later in the
analysis.
 
As a first step, a relation between the HI line area and the FIR surface brightness is
obtained. The relation depends on the gas-to-dust ratio, grain properties, and
the radiation field illuminating the interstellar medium (ISM) along the
line-of-sight. No significant variations have been observed in the gas-to-dust
ratio apart from those associated with large scale metallicity variations.
Similarly, because of the diffuse nature of the HI clouds, no small scale
changes in the intrinsic dust properties or dust temperature are expected.
Under these conditions the FIR signal should have a linear dependence on the
HI column density.  Because each field is considered individually, possible
differences in the HI--FIR relation towards different regions can and will be
taken into account. For each field, an extrapolation to zero HI intensity
eliminates emission associated with the neutral ISM (for details, see
Sect.~\ref{Sect:cirrus_HI}). The remaining signal is equal to the sum of the
zodiacal light (ZL) and the CIRB. These components are not removed because
they are uncorrelated with the HI emission. Furthermore, the ZL has a smooth
distribution and remains practically constant within each of the areas covered
by individual ISOPHOT maps (see \'Abrah\'am et al. \cite{Abraham_1997}). If the ZL
level is known, the absolute value of the CIRB can be obtained. The ZL
estimation is described in detail in Sect.~\ref{sect:ZL}.

\section{Observations} \label{sect:obs}

We
study three low surface brightness fields that are labelled NGP, EBL22, and
EBL26. The field NGP is located at the North Galactic Pole, the field EBL22 is
similarly at a high ecliptic latitude, while the third one, EBL26, lies close
to the ecliptic plane (see Table~\ref{table:fields}). 
EBL26 was selected as a field with high ZL level with the purpose of
estimating the ZL contribution at the different wavelengths observed in this
project.

The observations of the hydrogen 21\,cm line were made with the Effelsberg
radio telescope in May 2002. The telescope beam has a FWHM of 9 arcminutes.
The areas mapped with the ISOPHOT instrument were covered with pointings at
steps of FWHM/2. 
The stray radiation was removed with a program developed by P. Kalberla (see
Kalberla \cite{Kalberla1982}, Hartmann et al. \cite{Hartmann1996}, Kalberla et
al. \cite{Kalberla2005}).

For details of the observations of the EBL fields and the associated data
reduction, see Appendix~\ref{app:obs}. The principles of ISOPHOT data
reduction and calibration of surface brightness measurements are explained in
Appendix~\ref{sect:techcal}.

\begin{table}
\caption{
Parameters of linear fits of FIR surface brightness versus the HI line area.
The 1-$\sigma$ error estimates determined with the bootstrap method are given
in parentheses. For NGP, the results correspond to a fit to the combined data
of the northern and southern sub-fields (see Appendix
\ref{sect:isophot_reduction}).
}
\begin{tabular}{lccc}
\hline \hline
Field   &   $\lambda$   &    Offset      &   Slope    \\
        &   ($\mu$m)    &   (MJy\,sr$^{-1}$) & ($10^{-3}$
        MJy\,sr$^{-1}$\,K$^{-1}$\,km$^{-1}$\,s) \\
\hline
 EBL22   &   90  & 5.53 (0.30)  & 35.15 (4.26) \\ 
 EBL22   &  150  & 3.56 (0.34)  & 38.91 (4.38) \\ 
 EBL22   &  180  & 3.10 (0.38)  & 27.95 (5.42) \\ 
 EBL26$^1$   &   90  & 18.66 (1.22)  & 17.28 (6.57) \\ 
 EBL26$^1$   &  150  & 6.38 (1.33)  & 27.42 (7.36) \\ 
 EBL26$^1$   &  180  & 6.00 (1.30)  & 25.76 (7.41) \\ 
    NGP &   90  & 6.31 (0.17)  & 24.60 (3.98) \\ 
    NGP &  150  & 3.53 (0.23)  & 28.61 (5.58) \\ 
    NGP &  180  & 3.20 (0.19)  & 31.91 (4.22) \\ 
\hline
\end{tabular}

$^1$Fit to data with $W(HI)<$200\,K\,km\,s$^{-1}$ only.
\label{table:fit}
\end{table}

\section{Analysis and results}

\subsection{Subtraction of Galactic cirrus emission using HI data}
\label{Sect:cirrus_HI}

The FIR surface brightness was correlated at each observed wavelength with the
integrated line area of the HI spectra. At each observed HI position the
average FIR signal was calculated using spatial weighting with a gaussian with
FWHM equal to 9$\arcmin$. Only those pointings are used where the centre of
the Effelsberg beam falls inside the FIR map. In addition to the observational
uncertainties, each data point was weighted in direct proportion to the 
fraction of the HI FWHM beam that was covered by FIR observations. Therefore,
the data close to FIR map boundaries get lower weight in the following
analysis.

\begin{figure*} 
\resizebox{!}{8.5cm}{\includegraphics{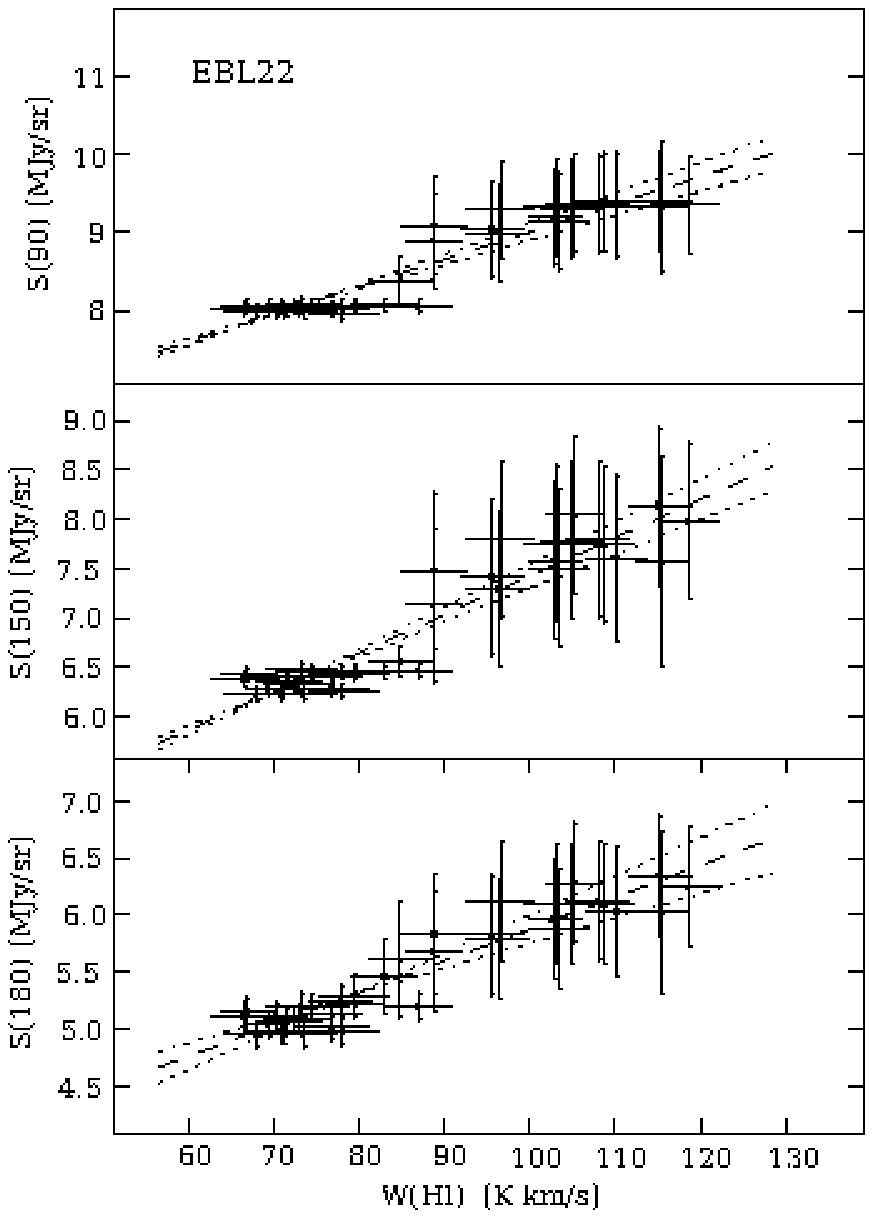}}
\resizebox{!}{8.5cm}{\includegraphics{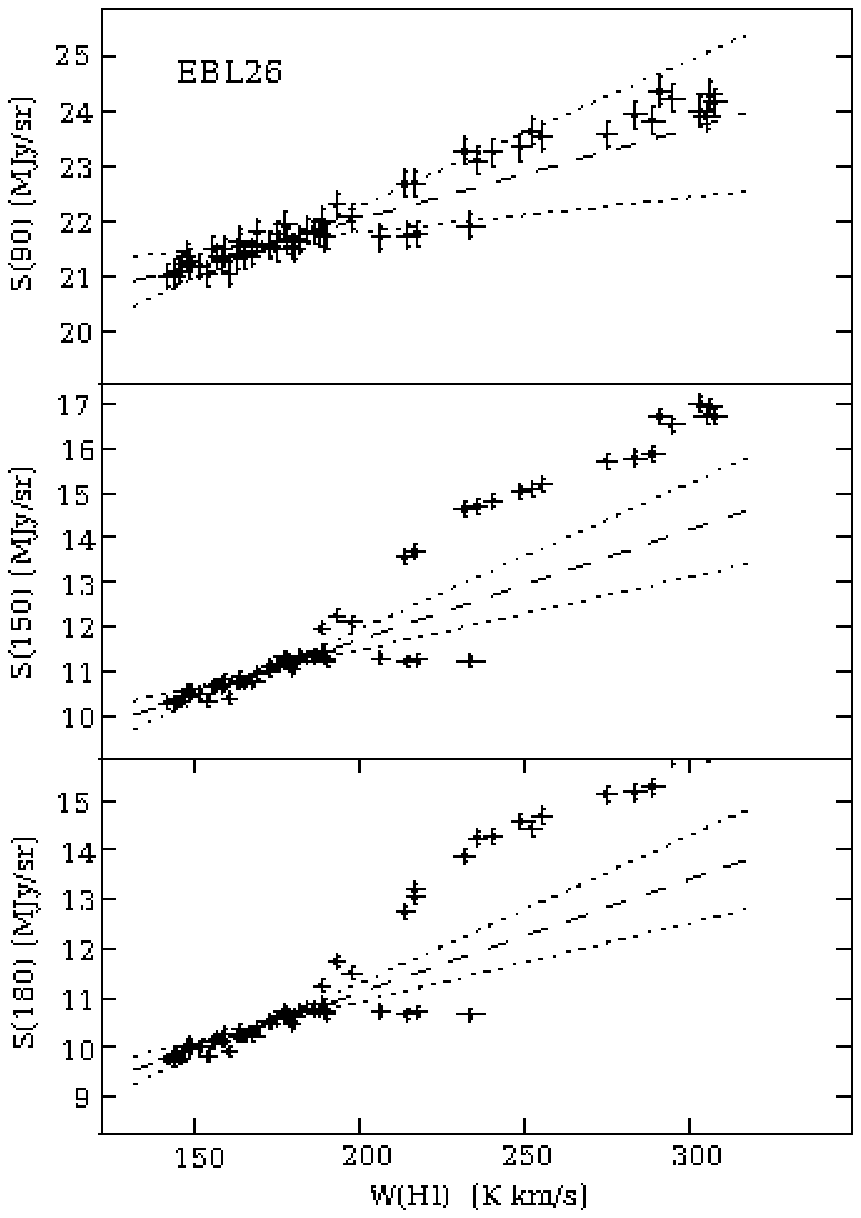}}
\resizebox{!}{8.5cm}{\includegraphics{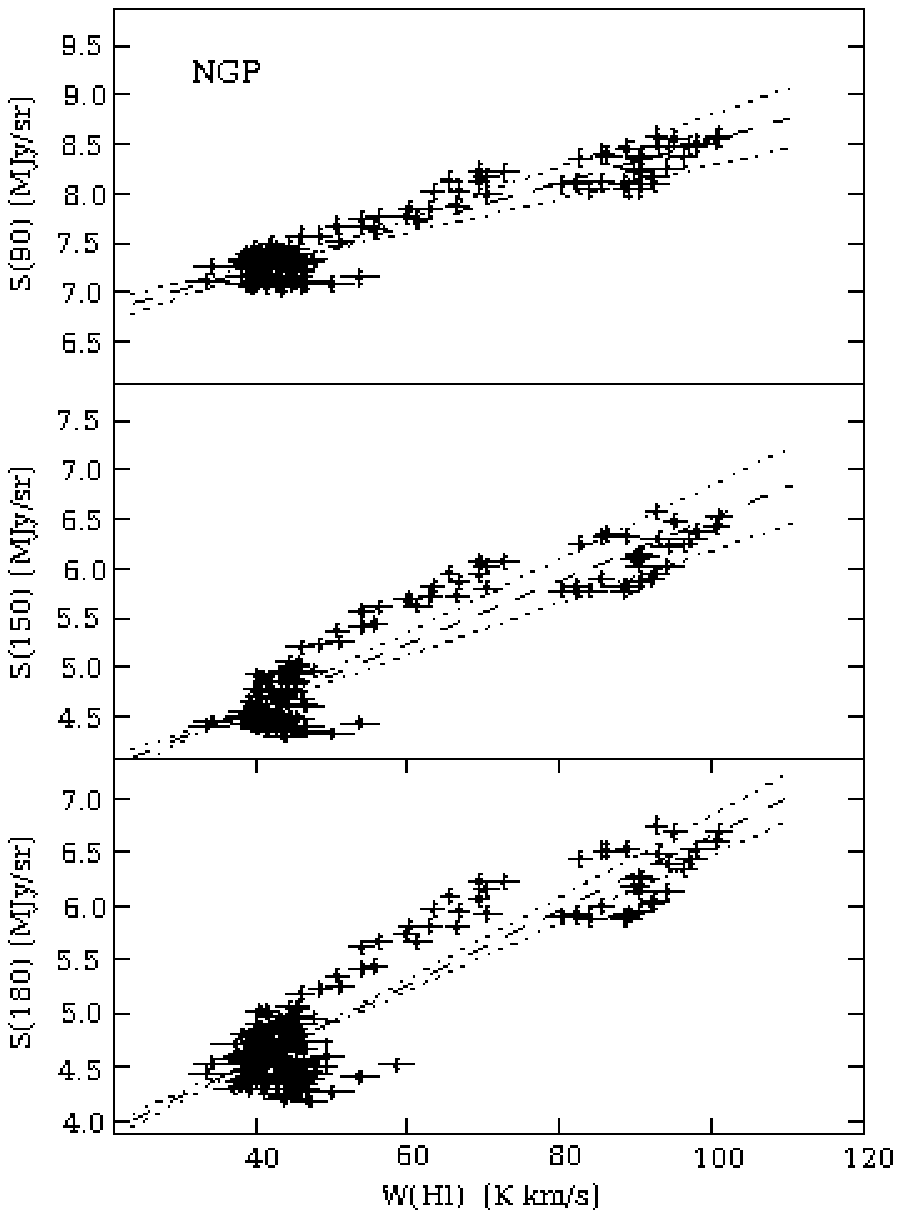}}
\caption{
FIR surface brightness as a function of HI line area $W$(HI) in the three EBL
fields, EBL22 ({\em left}), EBL26 ({\em middle}), and NGP ({\em right}). Each point
corresponds to one pointing of the HI observations. The uncertainties in the
HI line area are estimated based on the noise in velocity channels outside
detected HI emission. For each HI spectrum the corresponding average FIR
signal has been calculated using for weighting a gaussian with FWHM=9$\arcmin$. The
corresponding error bars are based on error estimates reported by PIA from
which the formal uncertainties of the weighted mean are calculated. The long
dashed line shows the result of a linear fit that takes into account the
uncertainties in both variables. The dotted lines indicate 67\% confidence
intervals that are obtained with the bootstrap method.
} 
\label{fig:HI_corr}
\end{figure*}

The obtained correlations are shown in Fig.~\ref{fig:HI_corr}. For FIR
observations the plotted error bars are based on the statistical uncertainties
reported by the PIA. 

The figures include linear fits that take into account the estimated
uncertainties in both FIR and HI data. 
The slopes and zero points of the fit are given in Table~\ref{table:fit}.
In field EBL26 there is a clear break
in the relation above $W$(HI)=200\,K\,km\,s$^{-1}$ that may indicate the
presence of molecular gas. There is also one fairly bright galaxy that is
located in the region of higher cirrus emission and may have affected the
correlation. Therefore, in the field EBL26 the linear fitting was carried out
using only data below $W(HI)=200$\,K\,km\,s$^{-1}$.  In the other fields the
hydrogen column densities are in general smaller, $W(HI) \la
100$\,K\,km\,s$^{-1}$, so that the fraction of molecular gas can be expected
to be insignificant.

The offsets thus obtained correspond to an extrapolation to zero HI column
density. To the extent to which the remaining contributions of ionised and
molecular gas can be ignored (see below), the values correspond to the sum of
CIRB and the zodiacal light.

\subsection{Subtraction of the zodiacal light} \label{sect:ZL}

The zodiacal light (ZL) emission is assumed to have a pure black body
spectrum. The colour temperature of the spectrum depends on the ecliptic coordinates
of the source and the solar elongation at the time of the observations.
Leinert et al. (\cite{Leinert2002}) have studied the variations of
mid-infrared ZL spectra over the sky using a set of observations made with the
ISOPHOT spectrometer. We use their results to fix the colour temperature of
the ZL spectra. 

The absolute intensity of the ZL emission in the FIR is estimated with the
help of shorter wavelength ISOPHOT observations made using the
ISOPHOT P detector in the absolute photometry observing mode PHT-05 (Laureijs
et al. \cite{Handbook}). Because the observations were made in regions of low
cirrus emission, the mid-infrared signal is completely dominated by the ZL.
The measurements were carried out close to the larger raster maps, in terms of
both time and position. Therefore, they give a good estimate for the zodiacal
light emission present in the raster maps. FIR absolute photometry
measurements were made at the same time and at the same positions. These are
used to make a correction for the contribution that the interstellar dust has,
conversely, on the measured mid-infrared values. The complete list of
observations is given in Table~\ref{table:ZL_observations}.

The derived ZL values obtained from the fits (ZL+cirrus) are listed in
Table~\ref{table:ZL_estimates}. The values are given at the nominal wavelengths
assuming a spectrum $\nu I_{\nu}=$constant. The uncertainties were
estimated based on the quality of the fits (see Appendix~\ref{sect:zlfit}). In
fields EBL26 and NGP, because error estimate of each of the two measurements is
itself uncertain, we conservatively take the average of the two error
estimates as the uncertainty of the mean.

\begin{table}
\caption{%
The estimated zodiacal light emission.  The columns are: (1) name of the
EBL field (see Appendix, Table~\ref{table:ZL_observations}), (2) 
temperature of the zodiacal light spectrum (Leinert et al.
\cite{Leinert2002}), (3)-(6) estimated intensity of the zodiacal light at
25\,$\mu$m, 90\,$\mu$m, 150\,$\mu$m, and 180\,$\mu$m, and (7) relative
uncertainty of the ZL value calculated on the basis of the difference of the fitted ZL
model and the observations in the range 7.3--90\,$\mu$m. The zodiacal light
estimates are given at the nominal wavelengths assuming a spectrum $\nu
I_{\nu}$=constant. For EBL26
and NGP, two separate positions were observed (see Fig.~\ref{fig:allsky} and
Table~\ref{table:ZL_observations}).
}

\begin{tabular}{lrrrrrrr}
\hline \hline
Field   &  $T_{\rm ZL}$ &   
        $I$(25)   &  $I$(90)  
        &  $I$(150)  &  $I$(180) &  rms \\
        &  (K)  &  \multicolumn{4}{c}{(MJy/sr)} &  error  \\
\hline
   EBL22      & 280  &  40.51 &   7.69 &   3.03 &   2.19 &   13\% \\ 
   EBL26\_ZL1 & 270  & 106.05 &  21.03 &   8.34 &   6.02 &   21\% \\ 
   EBL26\_ZL2 & 270  & 100.50 &  19.84 &   7.87 &   5.68 &   7\% \\ 
   EBL26 (aver.)      & 270  & 100.98 &  19.94 &   7.91 &   5.71 &   14\% \\
   NGP\_ZL1   & 260  &  28.65 &   5.95 &   2.38 &   1.72 &   5\% \\ 
   NGP\_ZL2   & 260  &  29.15 &   6.06 &   2.42 &   1.75 &   18\% \\ 
   NGP (aver.)       & 260  &  28.69 &   5.96 &   2.38 &   1.72 &   12\% \\
\hline
\end{tabular}

\label{table:ZL_estimates}
\end{table}

\subsection{Estimated CIRB levels and their uncertainties} \label{sect:estimates}

Table~\ref{table:EBL_estimates} lists the CIRB levels that are estimated based
on the linear fits between FIR and HI data (Table~\ref{table:fit}) and the
zodiacal light values of Table~\ref{table:ZL_estimates}. The uncertainties are
obtained by adding in quadrature the estimated errors of the offsets from
Table~\ref{table:fit}, the errors of the zodiacal light values from
Table~\ref{table:ZL_estimates}, and the error resulting from the dark current
subtraction (see Appendix~\ref{sect:calibration}),
\begin{equation}
\sigma_{\rm tot}^2  = \sigma_{\rm offset}^2  + \sigma_{\rm ZL}^2  +
\sigma_{\rm DC}^2.
\end{equation}
The uncertainty due to the dark current is estimated to be $\sigma_{\rm
DC}=$0.25-0.30\,MJy\,sr$^{-1}$ and it is likely to be the main factor
affecting the uncertainty of the zero point of the FIR observations (see
Appendix~\ref{sect:calibration}). 

The results obtained for the three individual fields can be combined, deriving our
final estimates for the CIRB and its uncertainty. 
In
the case of the field EBL26 the values are relatively unprecise because of the
high ZL level. This uncertainty is reflected in the error
estimates. 
Combining the results we get average values
$-0.54\pm$0.65\,MJy\,sr$^{-1}$, 0.83$\pm$0.41\,MJy\,sr$^{-1}$,
1.26$\pm$0.37\,MJy\,sr$^{-1}$,
at 90\,$\mu$m, 150\,$\mu$m, and 180\,$\mu$m, respectively, as given in the
last line of Table~\ref{table:EBL_estimates}.

The 90\,$\mu$m values are very low, because in both the EBL22 and EBL26 fields
negative values are obtained.  In the case of EBL26 the negative value is not
surprising, because the expected CIRB level is only a small fraction of the
zodiacal light which itself has a considerable statistical uncertainty.
Therefore, the result is sensitive also to any systematic errors of the ZL
estimates. Apart from the results at 90\,$\mu$m, the variation between fields
is only slightly larger than expected on the basis of the quoted error
estimates. At 150\,$\mu$m a negative value is obtained for EBL26 which,
nevertheless, is less than 2-$\sigma$ below the highest values. 

At 90\,$\mu$m we can derive for EBL only an upper limit. The 150\,$\mu$m and
180\,$\mu$m bands are close to each other and the CIRB values should be very
similar.  Therefore, based on the three fields and the two frequency bands, we
can calculate, as a weighted average, an estimate for the CIRB in the range
150--180\,$\mu$m. The result is
1.08$\pm$0.32\,MJy\,sr$^{-1}$. The result would not change significantly (less
than $1-\sigma$) even if either EBL22 or EBL26 were omitted from the
analysis.

\begin{table}
\caption{
Estimated level of the CIRB for the individual fields. The values correspond
to the difference between the offsets listed in Table~\ref{table:fit} and the
ZL estimates of Table~\ref{table:ZL_estimates}. No colour correction was
applied. The 1-$\sigma$ estimated statistical errors are given in parentheses
(see text).
}
\begin{tabular}{lrrr}
\hline \hline
Field   &   $I$(90$\mu$m) & $I$(150$\mu$m) & $I$(180$\mu$m) \\
        &   (MJy\,sr$^{-1}$) &   (MJy\,sr$^{-1}$) &   (MJy\,sr$^{-1}$) \\
\hline \hline
EBL22       &   -2.16 (1.04)  &   0.53 (0.52) &   0.91 (0.47)  \\ 
EBL26$^1$   &   -1.28 (3.05)  &  -1.53 (1.73) &   0.29 (1.53)  \\ 
NGP         &    0.35 (0.74)  &   1.15 (0.37) &   1.48 (0.28)  \\ 
average     &   -0.54 (0.65)  &   0.83 (0.41) &   1.26 (0.37)  \\
\hline
\end{tabular}
\\$^1$Fit to data with $W(HI)<$200\,K\,km\,s$^{-1}$ only.

\label{table:EBL_estimates}
\end{table}

\subsection{The reliability of the CIRB values} \label{sect:reliability}

In addition to the statistical uncertainties, the results are affected by
systematic errors. The CIRB estimates are not affected by the HI antenna
temperature scale. However, the presence of unsubtracted stray radiation could
affect the HI zero point of the HI data and, thus, lower the CIRB values. We
cannot directly estimate the presence of residual stray radiation in the HI
data. However, in Sect.~\ref{sect:HImeas} we compare some of our HI
spectra with data from the Leiden/Dwingeloo survey (Hartmann \& Burton
\cite{Hartmann1997}; Kalberla et al. \cite{Kalberla2005})
and we find that the residual stray radiation is likely to be less than
4\,K\,km\,s$^{-1}$ which, assuming a slope of 29$\times
10^{-3}$\,MJy\,sr$^{-1}$\, (K\,km\,s$^{-1}$)$^{-1}$ (see
Table~\ref{table:fit}), corresponds to 0.12\,MJy\,sr$^{-1}$. In this case HI
stray radiation would not be a major source of error.

The relative calibration of the ISOPHOT FIR cameras and the P-detectors
directly affects the estimated FIR ZL levels and is probably the most
important source of systematic errors. According to the ISOPHOT Handbook
(Laureijs et al. \cite{Handbook}) the absolute accuracy of the C100 and C200
cameras and the P-detectors is typically of the order of 20\%. If there were a
difference of 10\% in the {\em relative} calibration of the MIR and FIR bands,
this would cause a similar percentual error to the ZL estimates.  The fact
that we obtained negative CIRB values at 90\,$\mu$m, especially when the
absolute level of the ZL is high, suggests that the FIR ZL levels may have
been overestimated. The effect of an error of 10\% would range from
$\sim$2\,MJy\,sr$^{-1}$ at 90\,$\mu$m in the field EBL26 to
$\sim$0.2\,MJy\,sr$^{-1}$ at 150--180\,$\mu$m in the field NGP. Taking into
account the relative weighting of the three fields, a 10\% error in ZL
corresponds to an uncertainty of 0.3\,MJy\,sr$^{-1}$ in the 150-180\,$\mu$m
CIRB estimate. Assuming a systematic uncertainty of this magnitude, the CIRB
estimate can be written as 1.08$\pm$0.32$\pm$0.30\,MJy\,sr$^{-1}$ where the
first error estimate refers to statistical and the second to systematic
uncertainties.

At 90\,$\mu$m the negative value obtained for EBL26 carries very little weight
and an additional 10\% systematic uncertainty in the ZL would correspond to an
additional uncertainty of 2\,MJy\,sr$^{-1}$. In EBL22 the CIRB values was
-2.16$\pm$1.04 and a 10\% systematic error in the ZL values would correspond
to 0.77\,MJy\,sr$^{-1}$. The CIRB value is $\sim$2-$\sigma$ below zero and
suggests that the ZL values may contain a systematic error of $\sim$10--20\%
that has reduced the obtained CIRB values. In the field NGP the 90\,$\mu$m
CIRB estimate was
0.35$\pm$0.74\,MJy\,sr$^{-1}$. Assuming a 10\% systematic uncertainty in the
ZL and adding the error estimates in quadrature, the CIRB estimate becomes
0.35\,$\pm$0.95\,MJy\,$sr^{-1}$ and we obtain a 2-$\sigma$ upper limit of
2.3\,MJy\,sr$^{-1}$.

\section{Discussion}

\subsection{Dust emission associated with ionised gas}

Our analysis was based on the correlation of HI emission and FIR intensity. 
So far we have omitted the possible effect that dust mixed with ionised gas
might have. The ionised component can affect the results only as far as it is
uncorrelated with the HI emission. 

Lagache et al. (\cite{Lagache2000}) decomposed the DIRBE FIR intensity into
components correlated with the neutral and the ionised medium.  The column density of
ionised hydrogen, $N(H^{+})$,
was estimated based on the
$H_{\alpha}$ line. 
They found that the infrared emissivity of dust associated with the ionised medium would
be very similar to the emissivity of dust within the neutral medium. However,
Odegard et al. (\cite{Odegard2007}) recently re-examined this issue, obtaining
significantly lower emissivity values for the ionised medium. The derived
2-$\sigma$ upper limit for the 100\,$\mu$m emissivity per hydrogen ion was
typically only 40\% of the emissivity in the neutral atomic medium.

We use the all-sky H$\alpha$ map produced by Finkbeiner
(\cite{Finkbeiner2003}) to examine the possible contribution of the ionised
medium to the FIR emission. The resolution of the $H_{\alpha}$ data is
6$\arcmin$ for fields EBL22 and EBL26 and one degree at the location of the
field NGP. The average $H_{\alpha}$ emission in the EBL22, EBL26, and NGP
fields is $\sim$0.7\,R, $\sim$0.5\,R, and $\sim$0.6\,R, respectively. The
$H_{\alpha}$ background contains small scale structure that may be caused by
faint point sources, mainly stars. Therefore, the quoted $H_{\alpha}$ levels
are not caused by the diffuse ISM only. For example, in NGP the
$H_{\alpha}$ image is dominated by an unresolved ($\sim$ one degree) emission
peak at the centre of the field, the nature of which remains unknown. Apart
from this, the $H_{\alpha}$ background does not show any significant gradients
or correlation with the FIR emission. Therefore, we consider only the
effect on the average FIR signal. Using the Lagache et al.
(\cite{Lagache2000}) conversion factors an $H_{\alpha}$ signal of 0.6\,R would
correspond to $\sim$0.5\,MJy\,sr$^{-1}$ in the FIR. Therefore, the CIRB
values could be overestimated by a similar amount. However, adopting the
Odegard et al. (\cite{Odegard2007}) 1-$\sigma$ upper limits, the contribution
from the ionised medium should remain below 0.1\,MJy/sr. 
Furthermore, in our analysis we have correlated the FIR emission only with HI
while in the quoted studies the FIR signal was correlated simultaneously with
both HI and H$^{+}$. Therefore, since HI and H$^{+}$ are themselves
correlated, the correction to be applied to our results should be
correspondingly smaller. Therefore, we believe that the possible effects due
to the presence of an ionised medium are small compared with the other
uncertainties given above.

\subsection{Dust emission associated with molecular gas}

If molecular gas is present, the HI lines will underestimate the total column
density of gas and, because the fraction of molecular gas increases with column
density, the relation between FIR emission and the HI intensity becomes
steeper. Our fields have low column density and, therefore, the fraction of
molecular gas should be low.  Hydrogen molecules cannot survive in clouds with
visual extinction below $A_{\rm V}\sim 0.1^{\rm m}$ and, consequently, no
molecular gas should exist in clouds with column densities below $N(H)\sim
2\times 10^{20}$\,cm$^{-2}$. Arendt et al. (\cite{Arendt1998}) detected a
steepening in the FIR vs. HI relation which, however, in different regions
took place at different column densities.  The effect could start already
around $N(H)\sim2\times 10^{20}$\,cm$^{-2}$ which corresponds to an HI line
area of $W(HI)\sim100\,$K\,km\,s$^{-1}$. 
Kiss et al. (\cite{Kiss2003}) observed a change in the spatial power spectra
of FIR surface brightness around $N(H)\sim 10^{21}$\,cm$^{-2}$. This was
similarly interpreted as a sign of the transition between atomic and molecular
phases.
In the EBL fields, molecular emission could be significant only in the EBL26
region, where the slope between HI and FIR data appears to change at $W(HI)
\sim 200\,$K\,km\,s$^{-1}$ (see Fig.~\ref{fig:HI_corr}). Below this
limit there is a good, linear correlation between the FIR surface brightness
and HI line area which also shows that toward those positions the fraction of
molecular gas is low. In the EBL estimation only data below
$W(HI)=200\,$K\,km\,s$^{-1}$ were used.

\subsection{Comparison with earlier results}

The earlier CIRB results in the FIR range are all based on measurements of
the COBE satellite. We have derived our CIRB estimates using the ISOPHOT
measurements, without relying on the COBE data even in the determination of
the ZL levels. Therefore, our result is the first completely independent CIRB
estimate after the COBE detections. Table~\ref{table:comparison} in
Appendix~\ref{sect:dirbe_ebl} lists the existing CIRB estimates in the FIR
wavelength range.

In the range 150--180\,$\mu$m our value is consistent with the DIRBE results
at 140\,$\mu$m. According to Kiss et al. (\cite{Kiss2006}) the COBE/DIRBE and
ISOPHOT FIR surface brightness values agree to within $\sim$15\% and,
therefore, the differences in the surface brightness scales are likely to be
smaller than our statistical uncertainty.  In the ZL subtraction, the relative
calibration of the ISOPHOT-P detector and the FIR cameras could introduce a
systematic error that has a magnitude comparable to that of the statistical
uncertainty. The low, even negative CIRB estimates obtained at 90\,$\mu$m
suggest that this systematic error causes the ZL estimates to be $\sim$10\%
too large. Taking into account our statistical and systematic uncertainties at
150--180\,$\mu$m, we cannot exclude even the highest DIRBE estimates close to
1.5\,MJy\,$^{-1}$.

At 90\,$\mu$m our 2-$\sigma$ upper limit of 2.3\,MJy\,sr$^{-1}$ is consistent
with the existing DIRBE results.

Based on the above values, the galaxies resolved with ISO FIR observations
account for some 5\% of the total CIRB (e.g., Juvela et al.
\cite{Juvela00}; H\'eraudeau et al. \cite{Heraudeau04}; Lagache \& Dole
\cite{lagache_dole_01}; Kawara et al. \cite{Kawara04}). A stacking analysis of
Spitzer measurements (Dole et al. \cite{Dole2006}) showed that galaxies
detected at 24\,$\mu$m contribute some 0.7\,MJy\,sr$^{-1}$ to the 160\,$\mu$m
sky surface brightness. Therefore, the results from galaxy counts and
measurements of the absolute level of CIRB are converging, and probably more
than half of the sources responsible for the CIRB have already been identified.

\section{Conclusions}

For the ISOPHOT EBL project far-infrared raster maps were obtained in
selected low-cirrus regions. We have analysed these observations and, by
correlating the FIR surface brightness with HI line areas measured with the
Effelsberg radio telescope, we derive estimates for the cosmic infrared
background in the wavelength range 90--180\,$\mu$m. We determined the level
of ZL emission using shorter wavelength ISOPHOT observations, without
relying on a model of the spatial distribution of the ZL emission on the sky.
Therefore, our results are independent of the existing COBE results.

{\flushleft Based on this study we conclude the following:}
\begin{itemize}
\item At 90\,$\mu$m we derived a 
2-$\sigma$ upper limit of 2.3\,MJy\,sr$^{-1}$ for the CIRB.
\item In the range 150--180\,$\mu$m we obtained a CIRB value of
1.08$\pm$0.32$\pm$0.30\,MJy\,sr$^{-1}$, where we quote separately the 
estimated statistical and systematic uncertainties.
\item The accuracy of the results was determined mostly by the accuracy of the
zodiacal light estimates and the dark signal subtraction.
\item Assuming the latest emissivity values of dust associated with the
ionised medium, the uncertainty related to the presence of ionised medium was
small compared with the other sources of uncertainty.
\end{itemize}

\begin{acknowledgements}
We thank the anonymous referee for valuable comments.
This work was supported by the Academy of Finland grants no. 115056, 107701,
124620, and 119641. ISOPHOT and the Data Centre at MPIA, Heidelberg, were
funded by the DLR and the Max-Planck-Gesellschaft. We thank P. Kalberla for his
help in the planning of the HI measurements and for performing the stray
radiation correction of these observations.
\end{acknowledgements}

\Online

\appendix


\section{The principles of surface brightness observations with ISOPHOT: data
reduction and calibration} \label{sect:techcal}

The most detailed description of the ISOPHOT instrument, its observing modes
(so called Astronomical Observation Templates, AOTs) and the corresponding
data analysis and calibration steps is given in the ISOPHOT Handbook
(Laureijs et al. \cite{Handbook}). In the following we describe recent calibration 
techniques which are beyond the scope of the Handbook and which are essential
for the determination of the EBL surface brightness.

\subsection{Absolute surface brightness calibration of ISOPHOT observations}
\label{sect:abssurfacebrightnesscal}

ISOPHOT was absolutely calibrated against a flux grid of celestial point 
source standards consisting of stars, asteroids and planets, thus covering
a fair fraction of the entire dynamic flux range from $\approx$100\,mJy up
to about 1000\,Jy. Each detector aperture/pixel was individually calibrated 
against these standards. Therefore, the basic ISOPHOT calibration is in 
Jy\,pixel$^{-1}$.

In order to derive proper surface brightness values in MJy\,sr$^{-1}$, the 
solid angles of each detector aperture/pixel must be accurately known

\begin{equation}
B_{\rm \lambda} = \frac{f_{\rm psf}^{aper}(0,0)}{ \Omega^{aper}_{\rm eff}} 
                  \cdot F_{\rm \lambda},
\end{equation}

with $B_{\rm \lambda}$ being the surface brightness, $\Omega^{aper}_{\rm eff}$
the effective solid angle of the pixel/aperture, $F_{\rm \lambda}$ the total
flux of a celestial standard and $f_{\rm psf}^{aper}(0,0)$ the fraction of
the Point Spread Function contained in the pixel/aperture (i.e. the 
convolution of the PSF with the aperture response) when being centred at
position (0,0). Hence, $F_{\rm \lambda} \cdot f_{\rm psf}^{aper}(0,0)$ is the 
flux per pixel.
  
ISOPHOT's effective solid angles have been determined by 2D-scanning of a 
point source over the pixel/aperture in fine steps dx' and dy' and measuring
the resulting intensity at each measurement point $(x'_{\rm i},y'_{\rm j})$,
the footprint, taking into account a non-flat aperture/pixel response:

\begin{equation}
\Omega^{aper}_{\rm eff} = 
\sum_{i}^{A}\sum_{j}\,f_{\rm psf}^{aper}(x'_{\rm i},y'_{\rm j})\,dx'\,dy'
\end{equation}

\begin{figure}[ht!]
\centering
\includegraphics[clip=true,width=8.5cm,angle=90]{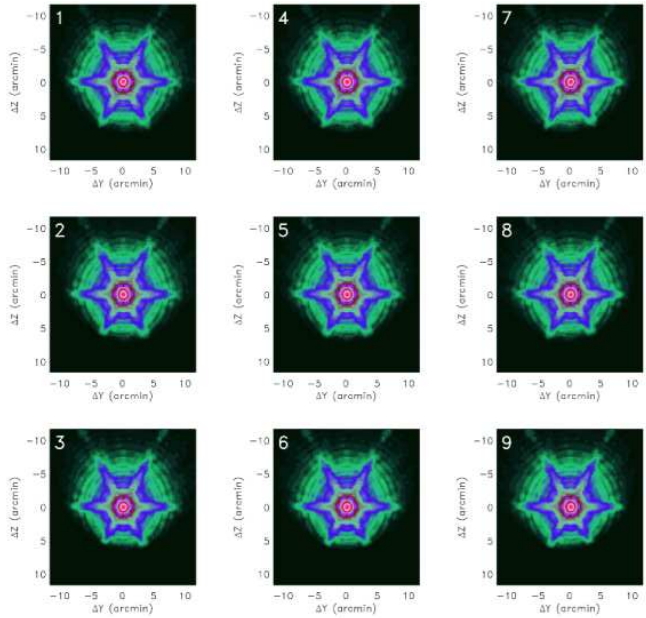}
\caption{Synthetic (outer part, i.e.\ green and blue coloured areas, modelled) 
         footprints (convolution of the ISO telescope PSF with the pixel 
         aperture response) of the 3 $\times$ 3 pixels of ISOPHOT's C100 array
         for the 60\,$\mu$m broad band filter. The solid angles of each pixel 
         are obtained by integration over the footprint area.}
\label{fi:c100_60umsyntheticfootprint}
\end{figure}

If the peak of the point source was located outside the aperture by 
$\sim$1/2 of the aperture size, the S/N of the resulting intensity dropped
so much that at this border the summation was complemented by a model of the 
broad band telescope PSF and adding up the corresponding PSF fractions 
$f_{\rm psf}^{aper}(x'_{\rm i},y'_{\rm j})$ out to $\pm$10\,arcmin assuming a 
flat response, but taking into account a cut by ISO's pyramidal central 
mirror feeding the 4 instrument beams. An example of such a synthetic 
footprint is shown in Fig.~\ref{fi:c100_60umsyntheticfootprint}.

The values of the solid angles used in PIA V11.3 are listed in 
Tables~\ref{tab:c100omega} and~\ref{tab:c200omega}.

\begin{table}[ht!]
\begin{center}
\caption{\label{tab:c100omega}
Effective solid angles for the 3 $\times$ 3 pixels of ISOPHOT's C100 array
for the 6 filters with central wavelengths $\lambda_{\rm c}$.}
\begin{tabular}{lcccc}
\hline \hline
\noalign{\smallskip}
filter  & $\lambda_{\rm c}$ & \multicolumn{3}{c}{$\Omega^{C100pixel\,i}_{\rm eff}$}~~~i = 1 $\ldots$ 9 \\
        &    ($\mu$m)       &\multicolumn{3}{c}{(10$^{-7}$ sr)} \\
\noalign{\smallskip}
\hline
\noalign{\smallskip}
C\_50  &  65 &  0.3737 &  0.4005 &  0.3929 \\
       &     &  0.4167 &  0.3963 &  0.4120 \\
       &     &  0.4064 &  0.3690 &  0.3943 \\
C\_60  &  60 &  0.3570 &  0.3925 &  0.3816 \\
       &     &  0.4087 &  0.3897 &  0.4044 \\
       &     &  0.3980 &  0.3608 &  0.3827 \\
C\_70  &  80 &  0.4252 &  0.4253 &  0.4276 \\
       &     &  0.4418 &  0.4177 &  0.4367 \\
       &     &  0.4326 &  0.3940 &  0.4299 \\
C\_100 & 100 &  0.4855 &  0.4517 &  0.4666 \\
       &     &  0.4679 &  0.4396 &  0.4632 \\
       &     &  0.4603 &  0.4194 &  0.4701 \\
C\_105 & 105 &  0.5006 &  0.4588 &  0.4764 \\
       &     &  0.4751 &  0.4454 &  0.4698 \\
       &     &  0.4678 &  0.4260 &  0.4804 \\
C\_90  &  90 &  0.4577 &  0.4403 &  0.4492 \\
       &     &  0.4568 &  0.4304 &  0.4519 \\
       &     &  0.4484 &  0.4086 &  0.4520 \\
\noalign{\smallskip}
\hline
\end{tabular}
\end{center}
\end{table}

\begin{table}[ht!]
\begin{center}
\caption{\label{tab:c200omega}
Effective solid angles for the 2 $\times$ 2 pixels of ISOPHOT's C200 array
for the 5 filters with central wavelengths $\lambda_{\rm c}$.}
\begin{tabular}{lccccc}
\hline
\hline
\noalign{\smallskip}
filter  & $\lambda_{\rm c}$ & \multicolumn{4}{c}{$\Omega^{C200pixel\,i}_{\rm eff}$}~~~i = 1 $\ldots$ 4 \\
        &    ($\mu$m)       &\multicolumn{4}{c}{(10$^{-7}$ sr)} \\
\noalign{\smallskip}
\hline
\noalign{\smallskip}
C\_160 & 170 &  1.782 &  1.940 &  1.895 &  1.781 \\
C\_200 & 200 &  1.810 &  1.996 &  1.988 &  1.878 \\
C\_180 & 180 &  1.792 &  1.960 &  1.927 &  1.815 \\
C\_135 & 150 &  1.759 &  1.898 &  1.825 &  1.711 \\
C\_120 & 120 &  1.722 &  1.843 &  1.708 &  1.587 \\
\noalign{\smallskip}
\hline
\end{tabular}
\end{center}
\end{table}

It should be noted that an absolute surface brightness calibration is more
accurate than an absolute calibration of a compact source of similar
brightness, since no background subtraction has to be performed, which
introduces an additional uncertainty. The accuracies quoted in the ISOPHOT
Handbook (Laureijs et al. \cite{Handbook}), Table 9.1 for extended emission 
take COBE/DIRBE photometry as the reference. By not referring to COBE/DIRBE 
photometry, the absolute surface brightness calibration for ISOPHOT's C100 
and C200 array is as good as that for bright compact sources, i.e.\ better 
than 15\%.

\subsection{New calibration products and strategies for PIA V11.3}

\begin{figure}[ht!]
\centering
\includegraphics[clip=true,width=8.5cm,angle=0]{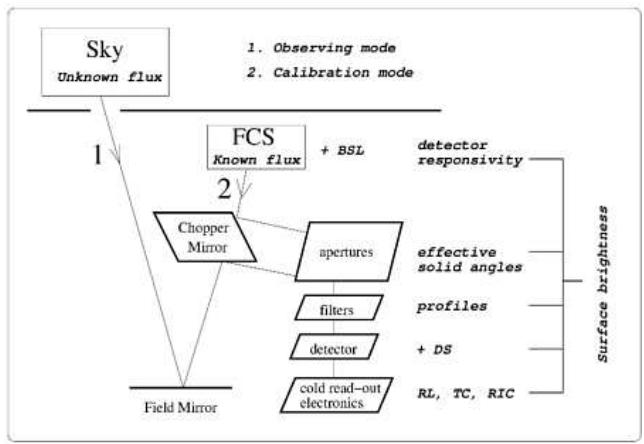}
\caption{Scheme of the ISOPHOT calibration steps associated with the different
         instrument components. The meaning of the abbreviations is the 
         following: BSL = Bypassing Sky Light correction, DS = detector Dark
         Signal, RL = Ramp Linearisation, TC = signal Transient Correction, 
         and RIC = Reset Interval Correction.}
\label{fi:phtcalibrationschema}
\end{figure}

For the very sensitive analysis needed for the EBL determination and, in
particular, an absolute surface brightness calibration that is as accurate as
possible, a number of calibration upgrades and new calibration features have
been developed and implemented in PIA V11.3. For the ones which are not
described in the ISOPHOT Handbook (Laureijs et al. \cite{Handbook}), we
provide a description and examples for the C100 detector in the following. An
overview of the individual calibration steps associated with different
instrument components is shown in Fig.~\ref{fi:phtcalibrationschema}. By
application of all these steps, instrumental artefacts are minimized, the
resulting detector signals are homogenized and a high calibration
reproducibility and accuracy is achieved.

\subsubsection{Detector responsivity calibration}

The absolute photometric calibration of an individual measurement is 
performed via a transfer calibration using the internal calibration sources.
This measures the actual responsivity of the detector and provides the 
absolute signal-to-flux conversion. It is a separate measurement of each 
observation mode by deflecting the chopper mirror to the field of view of the 
internal calibrator (Fine Calibration Source, FCS). The illumination level of 
the internal calibrator was not fixed but adjusted as much as possible to the 
expected brightness level of the sky. This was achieved by selecting an 
appropriate heating power for the internal source. There exists a calibration 
relation between this heating power and the optical power received by each 
detector pixel which is established from measurements on celestial standards.

\begin{figure*}[ht!]
\centering

\includegraphics[clip=true,width=5cm,angle=90]{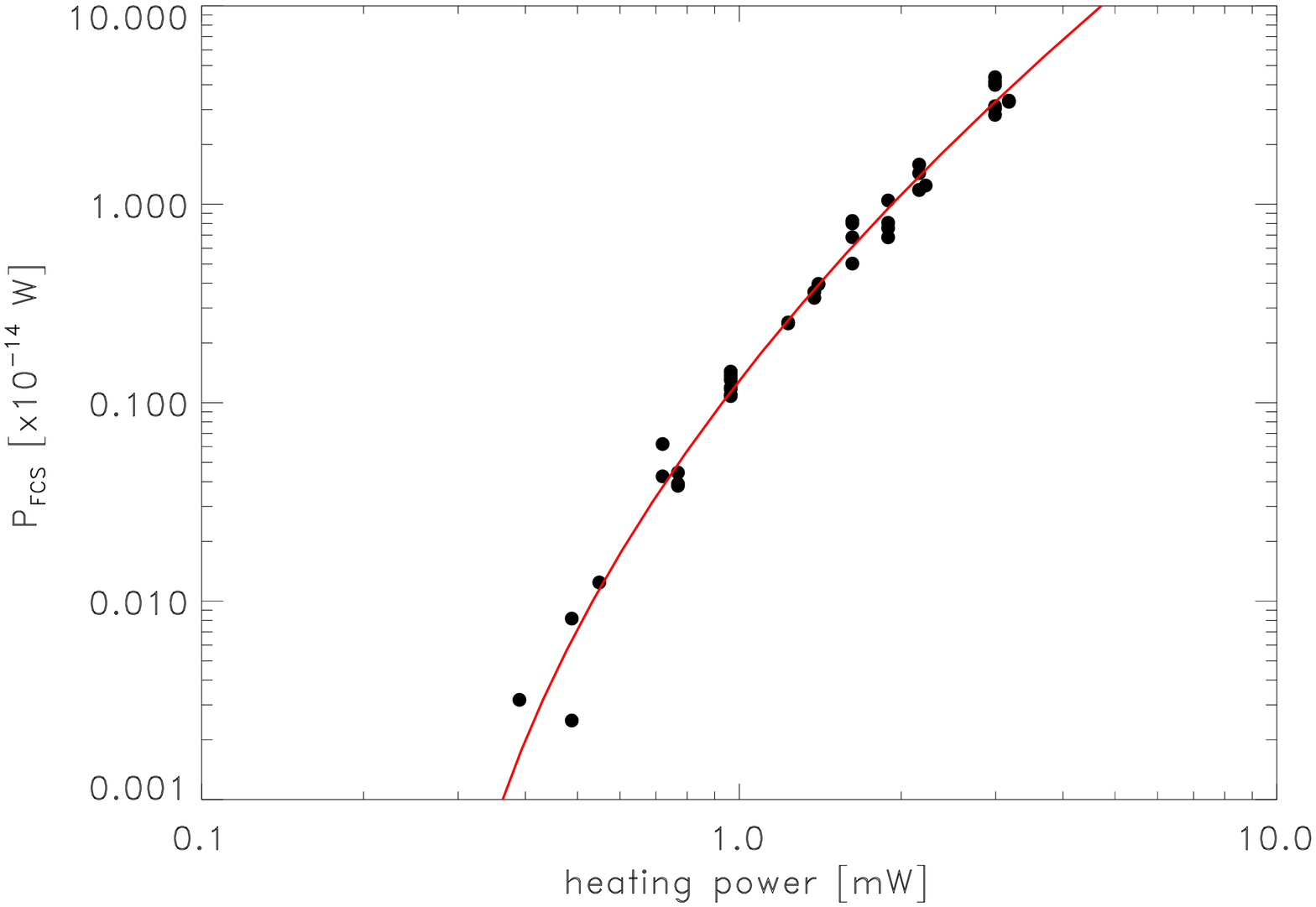}
\includegraphics[clip=true,height=5cm,angle=0]{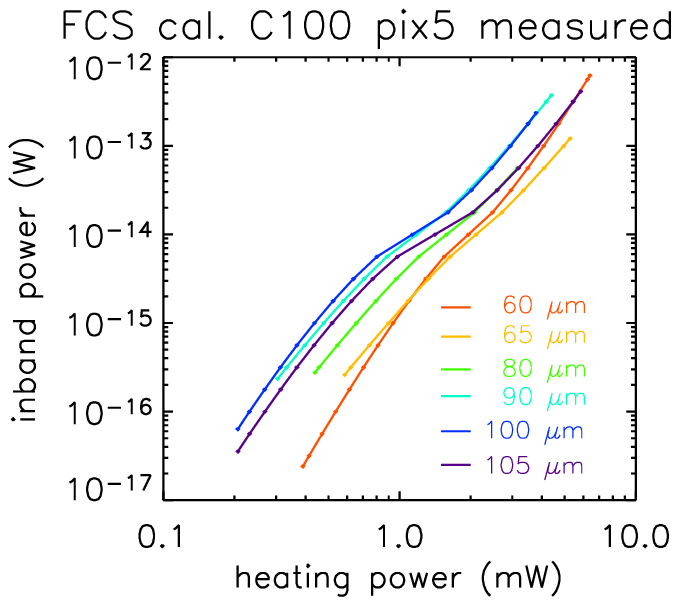}
\includegraphics[clip=true,height=5cm,angle=0]{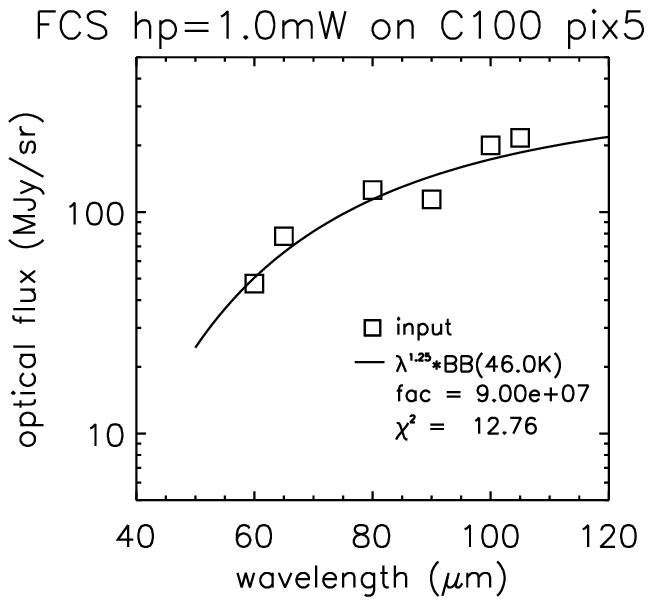}
\includegraphics[clip=true,height=5cm,angle=0]{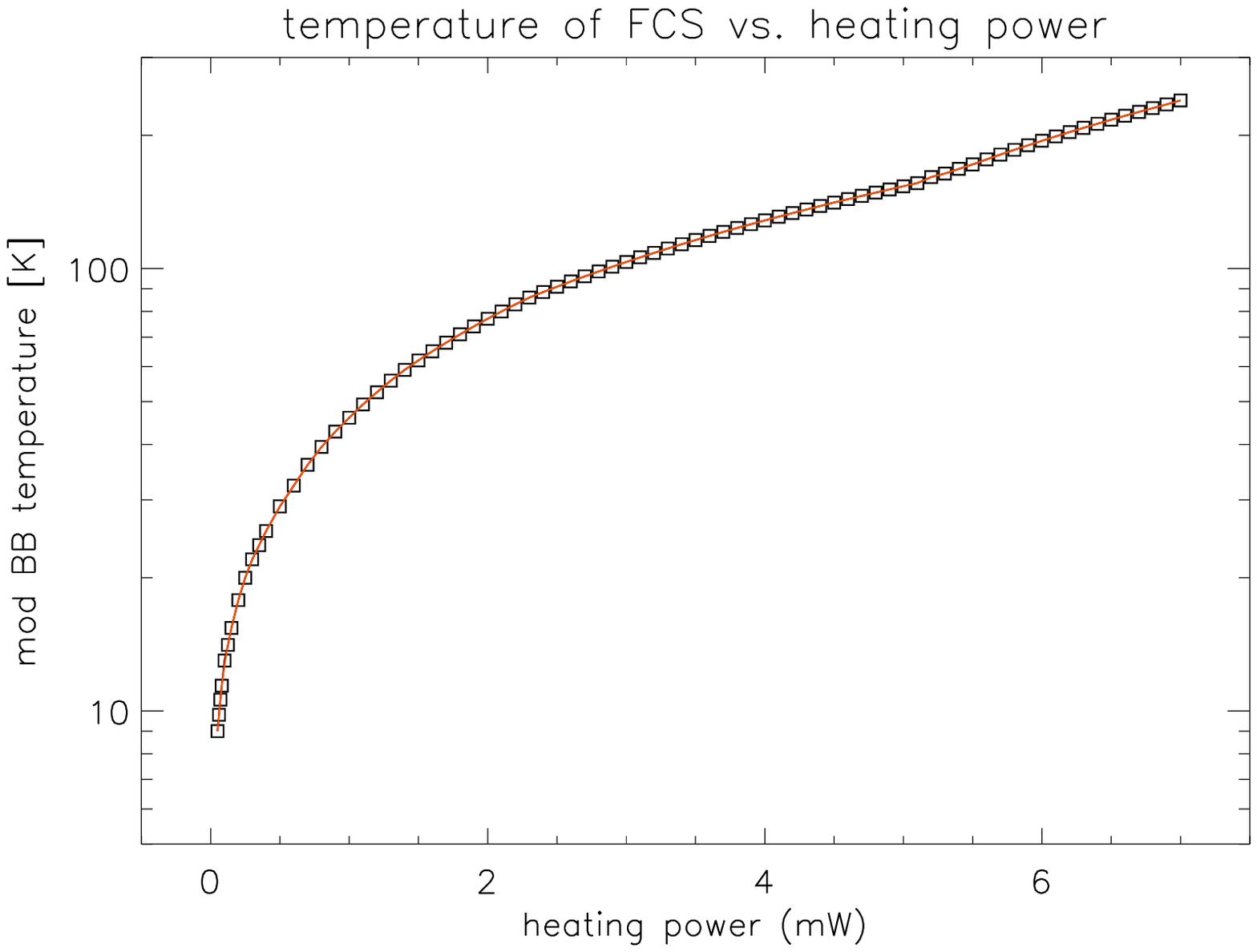}
\includegraphics[clip=true,width=5cm,angle=0]{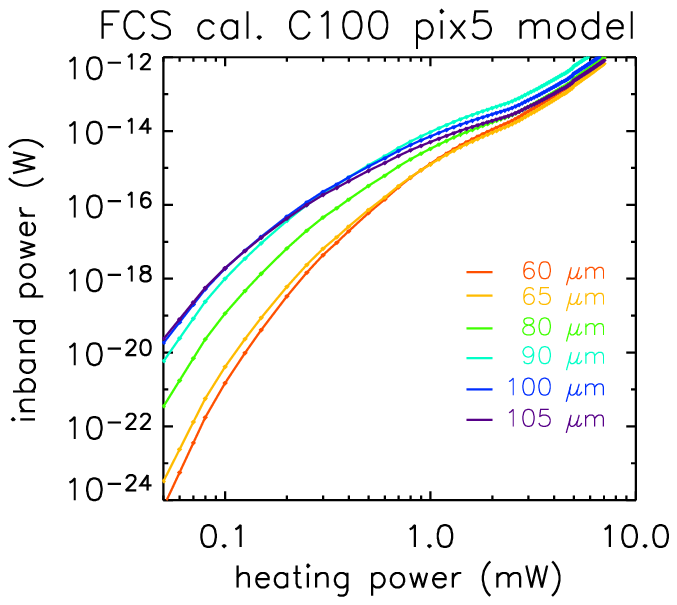}

\caption{Steps in the generation of a homogeneous and most complete 
         calibration of ISOPHOT's long wavelength internal calibration sources
         (FCS). This is illustrated for the central pixel (\#5) of the C100 
         array camera. {\it Upper left:} Measured relation between optical 
         power received on the detector and the heating power applied to the 
         internal source. Dots indicate the discrete measurements, the solid 
         line is a fit. {\it Upper right:} Display of the input curves for all
         C100 filters within the reliable heating power range. {\it Middle 
         left:} For a selected heating power (here: 1.0\,mW) monochromatic 
         and colour corrected fluxes of all filters are fitted by a modified 
         BB curve. {\it Middle right:} By repeating the fits with the same 
         modified BB type for the whole heating power range covered the 
         relation between heating power and temperature of the internal 
         source is established. {\it Lower centre:} By applying the FCS model
         the relation between optical power and heating power is homogenized
         and extended to the maximum heating power range covered by at least
         one measurement in any of the C100 or C200 filters.}
\label{fi:fcscalschema}
\end{figure*}

Therefore, for reliable and accurate transfer calibrations, the following 
requirements are put on the FCS:

\begin{itemize}
\item[1)] High reproducibility. This was better than 1\%, since the monitoring
          of the flux of faint standards was reproducible within a few percent,
          and this uncertainty was dominated by the signal noise 
          (Klaas et al. \cite{Klaas01}).
\item[2)] A very detailed characterization of the illuminated power depending
          on the heating power applied to the source. This is illustrated in
          Fig.~\ref{fi:fcscalschema}. It involves the following steps:
          \begin{itemize}
          \item[a)] For each C100 and C200 array filter all measurements of
                    celestial standards done in raster map mode were evaluated
                    such that for each pixel the background signal was 
                    properly subtracted and the resulting source signal was 
                    associated with the celestial standard flux. The ratio of 
                    the source signal and the simultaneously obtained FCS 
                    signal gave the illumination power by the FCS for the 
                    selected heating power. The discrete results were fitted 
                    and the reliable lower and upper heating power limits 
                    covered by measurements were identified 
                    (Fig.~\ref{fi:fcscalschema} upper left). The heating power
                    ranges were not identical or equally large for each filter
                    (Fig.~\ref{fi:fcscalschema} upper right). In general they
                    were shifted to smaller heating power values for longer
                    wavelengths.
          \item[b)] For fine discrete steps in heating power the inband powers
                    were read from the relations and were converted to 
                    monochromatic surface brightnesses by applying the 
                    bandpass conversions derived from the relative system 
                    response profiles, see ISOPHOT Handbook 
                    (Laureijs et al. \cite{Handbook}), section A.2,
                    and the solid angles of Tables~\ref{tab:c100omega} 
                    and~\ref{tab:c200omega}. These fluxes were fitted with
                    a modified BB curve after appropriate colour correction
                    (Fig.~\ref{fi:fcscalschema} middle left). If for a 
                    certain filter the selected heating power was outside
                    the reliable limits, the value of this filter was excluded
                    from the fit. The fit gave the temperature of the FCS for
                    the selected heating power. An additional constraint was
                    that the temperature had to be the same for the fit curves
                    of all pixels. C100 and C200 filter values had to be 
                    fitted independently because of the different detector 
                    areas and hence illumination factors, however, the fits 
                    were checked for consistent temperatures, because the 
                    illuminating FCS was the same for both detectors.
          \item[c)] This was achieved for the heating power range from 0.07
                    up to 6.5\,mW adopting an emissivity of the source 
                    $\propto \lambda^{1.25}$ yielding the temperature vs.\
                    heating power relation as shown in 
                    Fig.~\ref{fi:fcscalschema} middle right.
          \item[d)] By applying this FCS temperature model and the established
                    illumination factors for each pixel it was possible to
                    establish homogeneous calibration curves of the internal
                    reference source, thus polishing out measurement outliers
                    affecting the initial empirical curves. The multi-filter
                    approach connecting all curves and not treating them
                    individually enabled a large extension and a common range
                    for all filters: compare Fig.~\ref{fi:fcscalschema} lower
                    centre with Fig.~\ref{fi:fcscalschema} upper right. 
          \end{itemize}
\end{itemize}

\subsubsection{Bypassing sky light correction of FCS signal}

As a safety design against single point failures, ISOPHOT was not equipped 
with any cold shutter to suppress straylight when performing internal 
calibration measurements. Therefore, when deflecting the chopper onto the 
illuminated internal calibration (FCS) sources, some fraction of the power 
received on the detector did not come from the FCS but from sky light 
bypassing along non nominal light paths. Since this depends on the sky 
brightness it is subtracted in the transfer calibration measurements on 
celestial standards and hence has to be subtracted for any FCS measurement
in order to get a reproducible zero point. This was achieved by performing
a number of measurements on the switched-off, i.e.\ cold FCS, so that only 
the bypassing sky light contribution was measured.

\begin{figure}[ht!]
\centering
\includegraphics[clip=true,width=8.5cm,angle=0]{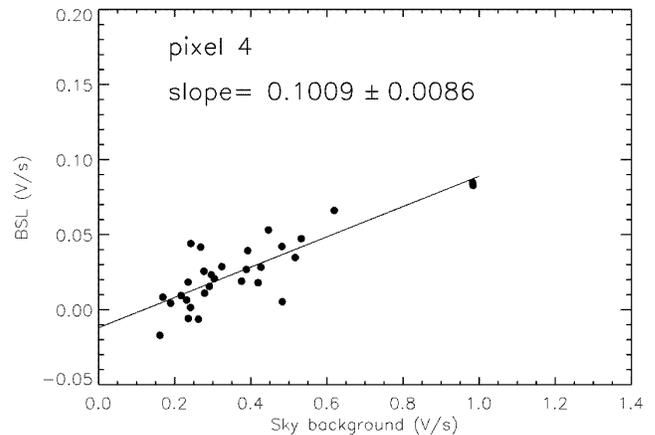}
\caption{Bypassing sky light contribution to the FCS signal depending on the 
         sky background.}
\label{fi:c100pix4bypassskylightcorr}
\end{figure}

The result for one C100 array pixel is shown in 
Fig.~\ref{fi:c100pix4bypassskylightcorr} which demonstrates a linear 
dependence of the bypassing sky light contribution to the FCS signal on the 
sky background. This correction was established for all C100 and C200 array
pixels. The bypassing sky light contribution contains the detector dark
signal contribution, cf. Sect.~\ref{sect:darksignal}. 

\subsubsection{Effective pixel/aperture solid angles}

These are described in the previous section~\ref{sect:abssurfacebrightnesscal}
and their values are compiled in Tables~\ref{tab:c100omega} 
and~\ref{tab:c200omega}.

\subsubsection{Filter profiles}

The bandpass system responses and the conversion factors from inband power
to a monochromatic flux, as well as colour correction factors are described
in the ISOPHOT Handbook (Laureijs et al. \cite{Handbook}).

\subsubsection{Detector dark signal}
\label{sect:darksignal}

\begin{figure}[ht!]
\centering
\includegraphics[clip=true,width=8.5cm,angle=0]{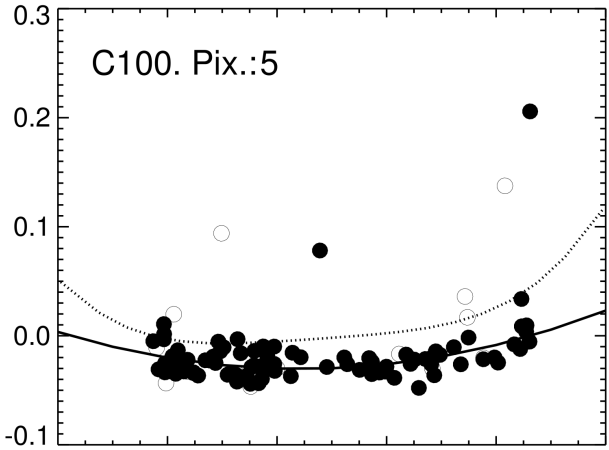}
\caption{Orbit dependent dark signal determination for the central pixel 5
         of ISOPHOT's C100 array. Dots represent individual measurements
         obtained during the entire ISO mission, filled and open signals
         identify a different reset interval in the integration of the dark
         signal. The solid line is the fit to the measurements providing
         the so-called default dark level. The dotted line is the default
         dark level of an older calibration version used before 2001.}
\label{fi:c100pix5darksignal}
\end{figure}

The detector dark signals were re-analyzed as described in
del Burgo (\cite{delBurgo02}). In this analysis special care was given to exclude those
dark measurements suffering from memory effects by preceding bright
illuminations, thus not representing the true dark level. An example of the
new results is shown in Fig.~\ref{fi:c100pix5darksignal} for the central pixel
5 of the C100 array. A slight orbital dependence is visible with an increase
of the dark signal towards the beginning and the end of the observational
window. It can also be noticed that there is a scatter of the dark
signals at the same orbit position and there are occasional large outliers.
These are not due to signal determination uncertainties, but are real
variations due to space weather effects on different revolutions over the ISO
mission.

\subsubsection{Ramp linearisation}

This was performed as described in the ISOPHOT Handbook (Laureijs et al.
\cite{Handbook}).
For ISOPHOT's far-infrared detectors two types of effects cause 
non-linearities of the integration ramps:
\begin{itemize}
\item[1)] De-biasing effects of the photoconductors operated with low bias 
          caused by feed-back from the integration capacitor.
\item[2)] Non-linearities generated in the cold read-out electronics.
\end{itemize}

\subsubsection{Signal dependence on reset interval correction}

\begin{figure}[ht!]
\centering
\includegraphics[clip=true,width=8.5cm,angle=0]{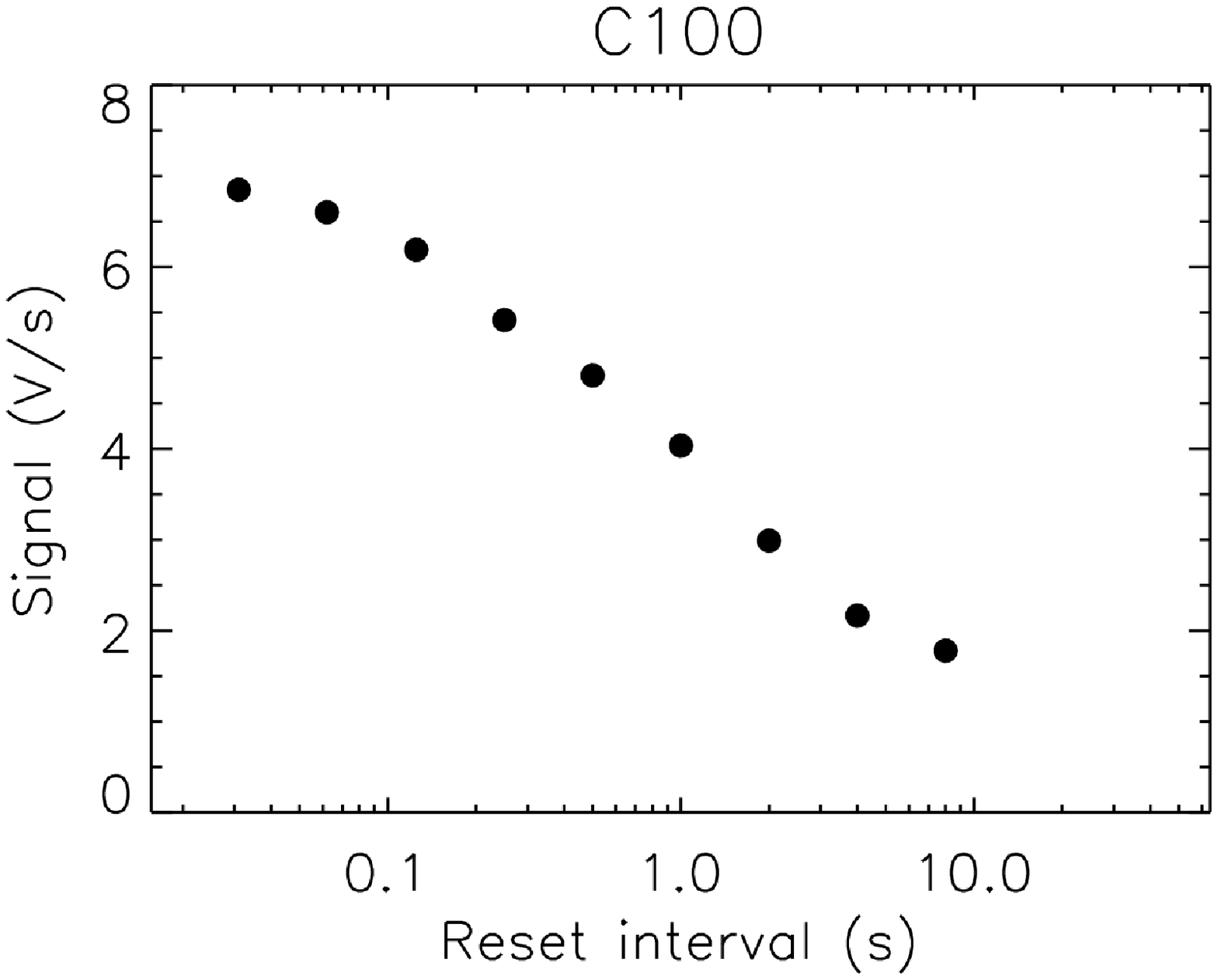}
\includegraphics[clip=true,width=8.5cm,angle=0]{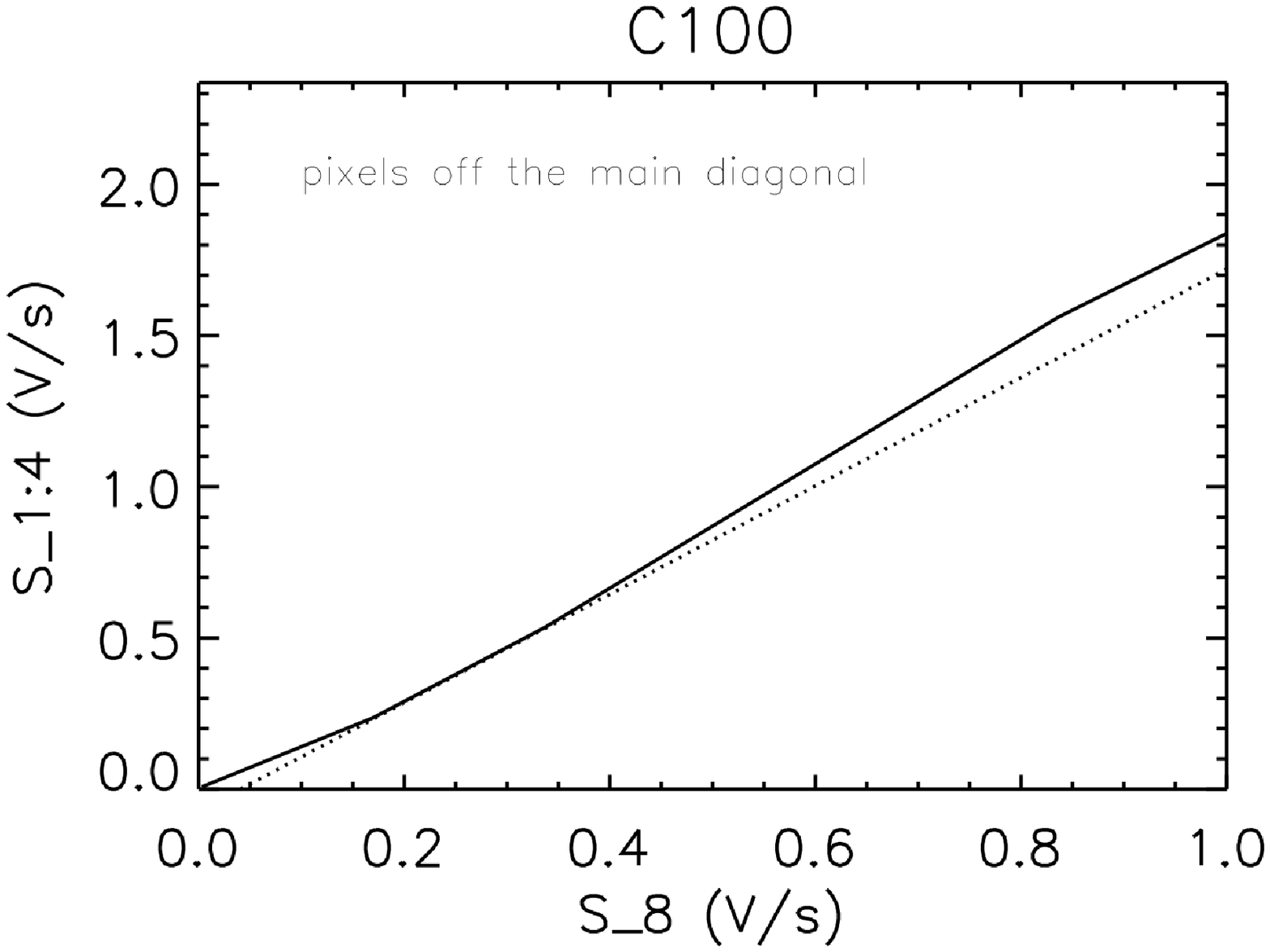}
\includegraphics[clip=true,width=8.5cm,angle=0]{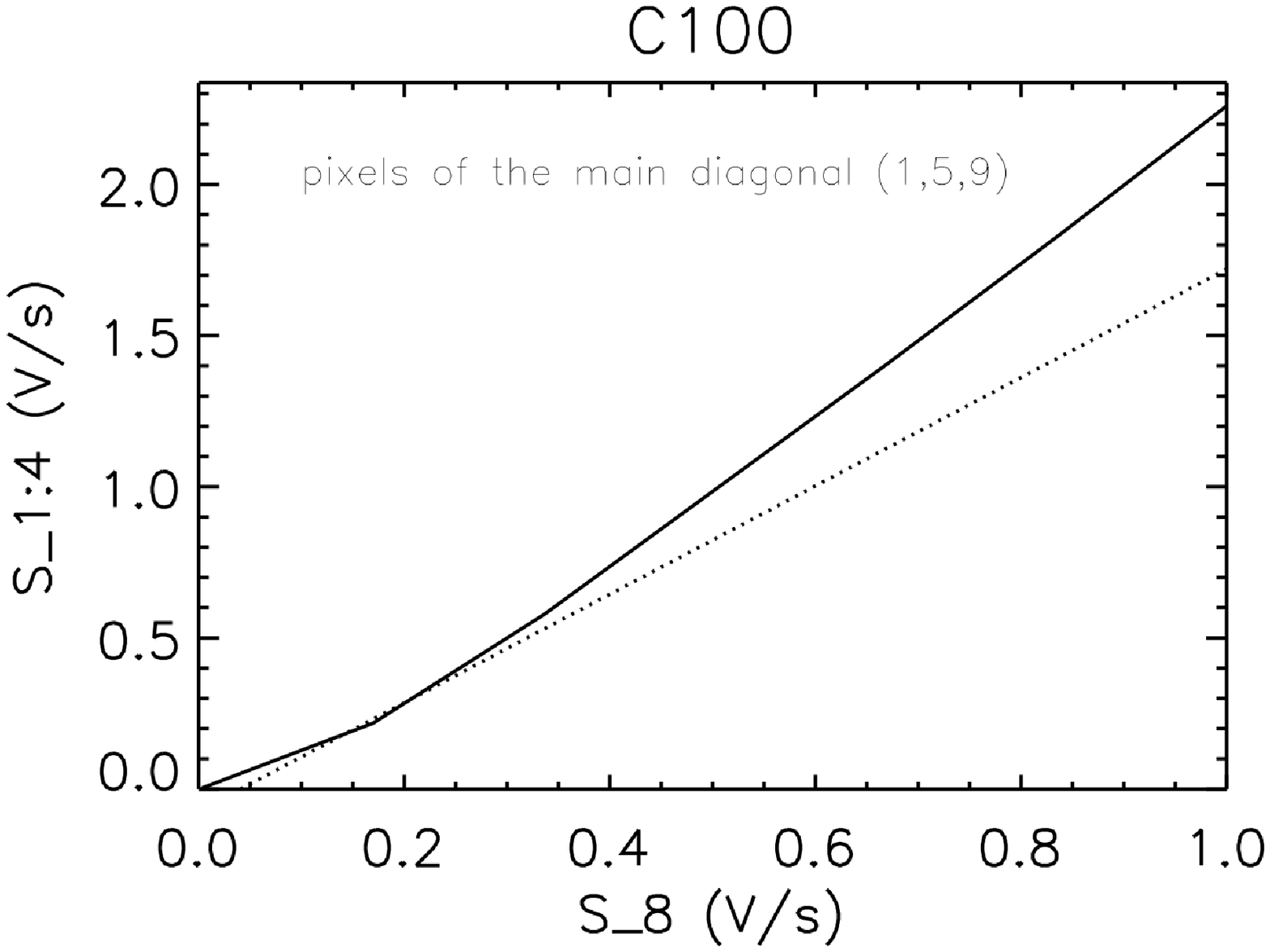}
\caption{Correction of the signal dependence on the selected reset interval. 
         {\it Upper panel:}
         Demonstration of the effect, showing the resulting signal versus
         the selected reset interval over the range from 1/32\,s up to 8\,s
         (reset intervals were commanded in powers of 2) under constant
         illumination. {\it Middle panel:} Solid line: Correction relation 
         for a reset interval of 8\,s w.r.t.\ the reference reset interval of 
         1/4\,s for all C100 array pixels, except the ones on the main 
         diagonal. Dotted line: old linear correlation used before the 
         re-analysis. {\it Lower panel:} Solid line: Correction relation for 
         a reset interval of 8\,s w.r.t.\ the reference reset interval of 
         1/4\,s for all C100 array pixels on the main diagonal (pixels \#1, 
         5, and 9). Dotted line: old linear correlation used before the 
         re-analysis (same as for middle panel).}
\label{fi:c100resetintervalcorr}
\end{figure}

Despite the ramp linearisation step, signals obtained under constant 
illumination, but with different reset intervals show a systematic difference,
see Fig.~\ref{fi:c100resetintervalcorr} upper panel. In order to have a
consistent signal handling of measurements with different reset interval 
settings applied - to optimize the dynamic range of the signal - all signals
were converted as if they were taken with a 1/4\,s reset interval. The
correction relations were established from special calibration measurements
applying the full suite of reset intervals under constant illumination and
this for different illumination levels. In this way signal corrections were
established for all reset intervals in the range 1/32\,s to 8\,s
(Fig.~\ref{fi:c100resetintervalcorr} middle and lower panel). While 
previously, as still described in the ISOPHOT Handbook (Laureijs et al.
\cite{Handbook}) a linear correlation with offset was used, a re-analysis
(del Burgo et al. \cite{delBurgo02}) yielded non-linear relations as shown in
Fig.~\ref{fi:c100resetintervalcorr}.  This latter analysis also found a
bi-modal behaviour for C100
array pixels, such that the pixels on the
main diagonal, \#1, 5 and 9, behaved differently from the rest of the pixels.
For the C200 array all pixels behaved in the same way.

\subsubsection{Signal transient correction}

\begin{figure}[ht!]
\centering
\includegraphics[clip=true,width=8.5cm,angle=0]{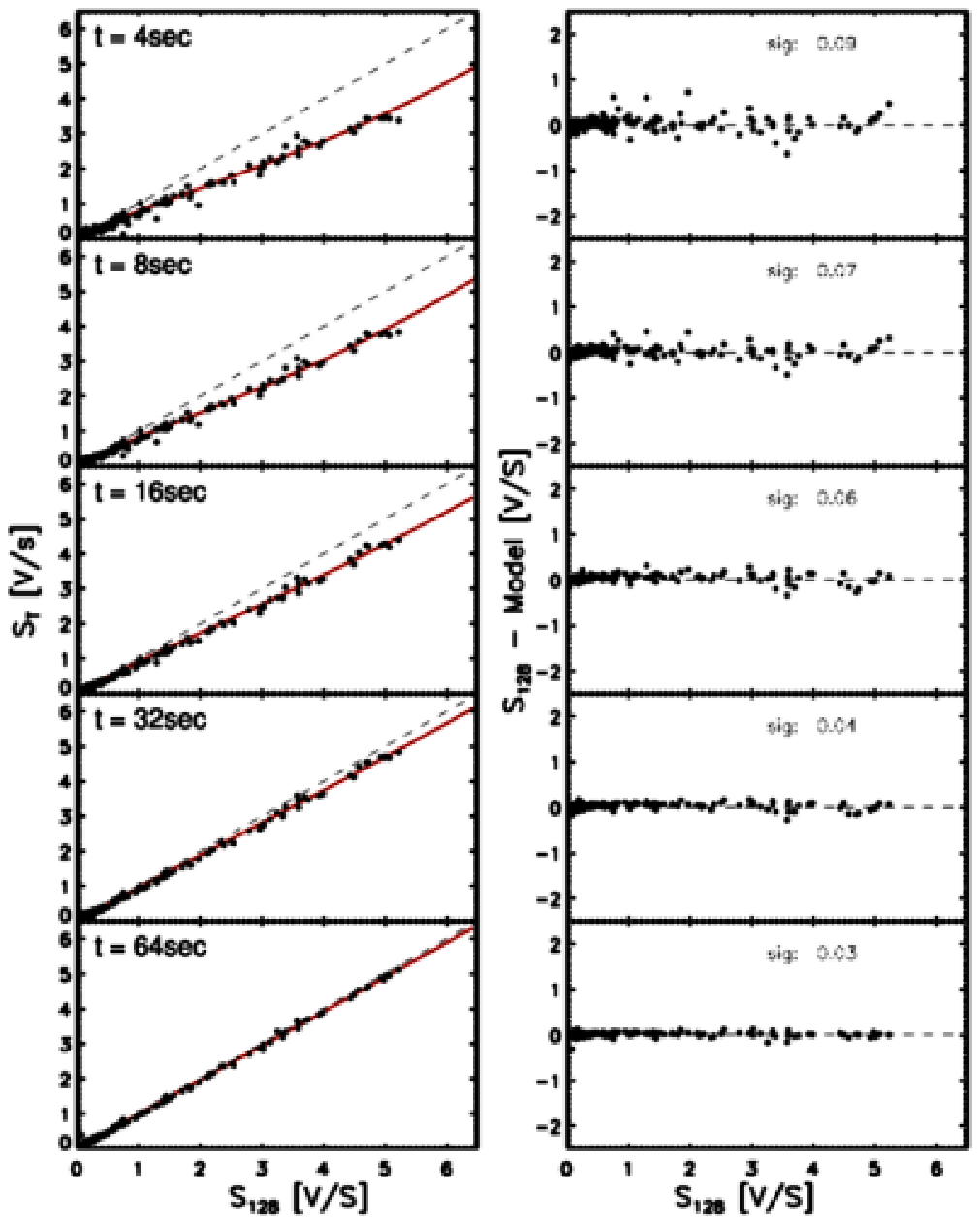}
\caption{Empirical signal transient correction for ISOPHOT's C100 array.
         The left column shows the signal loss for integration times of
         4, 8, 16, 32, and 64\,s (commendable integration times of ISOPHOT
         detectors) with regard to the reference time of 128\,s. The red line 
         is a fit through the measured points over the covered signal range 
         and is used as the correction relation. The right column shows the
         residuals after applying this correction.}
\label{fi:c100transientcorr}
\end{figure}

The ISOPHOT detectors were photoconductors operated under low background 
conditions provided by a cryogenically cooled spacecraft. Under these 
conditions they showed the behaviour that the output signal was not 
instantaneously adjusted to a flux change but rather, following an initial
jump by a certain fraction of the flux step, the signal adjusted with some 
time constant to the final level, see e.g. Acosta et al.
(\cite{Acosta-Pulido00}). In particular the ISOPHOT C100 detector showed
significant transient behaviour. This time constant depended on the detector
material (doping of the semi-conductor and its contacts), the flux step, the
direction of the flux step (dark to bright versus bright to dark) and the
illumination history. Attempts had been made to model this behaviour (Acosta
et al. \cite{Acosta-Pulido00}), but no unique description could be found for
the FIR detectors. To overcome this effect at least partly the method of
transient recognition was implemented in the ISOPHOT analysis software as
described in the ISOPHOT Handbook (Laureijs et al. \cite{Handbook}) using the
most stable part of the measurement for signal determination. Finally, another
approach was to use a data base of long measurements with signals stabilising
and to determine the deviation from the end level for shorter intermediate
times (del Burgo et al. \cite{delBurgo02}), see
Fig.~\ref{fi:c100transientcorr} for an illustration. A measurement time of
128\,s was used as reference, because
\begin{itemize}
\item[1)] Most calibration measurements in staring mode were performed with
          this basic measurement time.
\item[2)] In most cases the signals stabilised within this time.
\end{itemize}

For the C200 array pixels the signal transient effect is considerably smaller
and faster and therefore it is sufficient to apply the transient recognition
as described in the ISOPHOT Handbook (Laureijs et al. \cite{Handbook}).

\section{Observations and data reduction for the EBL fields} \label{app:obs}

\subsection{ISOPHOT observations}

The following tables give details of the ISOPHOT observations used in the
paper. Table~\ref{table:tdt} lists the raster maps and absolute photometry
measurements that were made at 90, 150, and 180$\mu$m.
Correspondingly, Table~\ref{table:ZL_observations} lists observations used for
the determination of the zodiacal light levels. These include both
mid-infrared measurements carried out with the ISOPHOT-P detector and longer
wavelength absolute photometry measurements carried out with the C100 and C200
cameras.

\begin{table}
\caption{List of ISOPHOT observations of EBL fields carried out in the
PHT-22 and PHT-25 observation modes.}
\begin{tabular}{lrrcccc}
\hline \hline
Field /  &  $\lambda$  &  RA        &  DEC        &  Size     &  TDT  \\
AOT              &   ($\mu$m)  &  (J2000)   &  (J2000)    &              &  number      \\
\hline

EBL22          &             &              &                 &              &              \\
   \, PHT22  &     90      &  2 26 34.4 & -25 53 49       &  32$\times$3 &  82101111  \\
   \, PHT22  &    150      &  $-''-$     &  $-''-$         &  $-''-$      &  81901910  \\
   \, PHT22  &    180      &  $-''-$     &  $-''-$         &  $-''-$      &  81901509  \\
   \, PHT22  &    90       &    2 32 47.9  &  -25 54  6.0   &    32$\times$1  &    82101115 \\
   \, PHT22  &   180       &    $-''-$   &  $-''-$    &    $\times$132  &    81901513 \\
   \, PHT22  &   150       &    $-''-$   &  $-''-$    &  $\times$1  32  &    81901914
   \\
   
   \, PHT25  &   150  &    2 26 34.4     &  -25 53 52.8    &     1        &    81901927 \\
   \, PHT25  &   180  &    2 26 34.4     &  -25 53 52.8    &     1        &    81901529 \\
   \, PHT25  &    90  &    2 26 34.4     &  -25 53 52.9    &     1        &    82101126 \\   
       
EBL26           &             &               &                &              &  \\
   \, PHT22  &     90      &  1 18 14.5   & 1 56 39       &  32$\times$4 &  58600319 \\
   \, PHT22  &    150      &  $-''-$       & $-''-$         &  $-''-$      &  58600418 \\
   \, PHT22  &    180      &  $-''-$       & $-''-$         &  $-''-$      &  58600517 \\
   \, PHT25  &   150       &    1 17 39.3  &    2  3  9.8   &     1  &    60302067 \\
   \, PHT25  &   180       &    $-''-$     &    $-''-$      &     1  &    60302275 \\
   \, PHT25  &    90       &    1 19 30.2  &    1 23 59.2   &     1  &    58600350 \\
   \, PHT25  &    90       &    $-''-$     &    $-''-$      &     1  &    58600342 \\
   \, PHT25  &   150       &    1 19 30.2  &    1 23 59.2   &     1  &    58600443 \\
   \, PHT25  &   150       &    $-''-$     &    $-''-$      &     1  &    58600451 \\
   \, PHT25  &   180       &    $-''-$     &    $-''-$      &     1  &    58600545 \\
   \, PHT25  &   180       &    $-''-$     &    $-''-$      &     1  &    58600553 \\
             
NGP(N)          &             &               &                &              &  \\
   \, PHT22  &     90      &  13 43 53.8   & 40 11 42      &  32$\times$4 &  55800211 \\
   \, PHT22  &    150      &  $-''-$       &  $-''-$        &  $-''-$      &  56200110 \\
   \, PHT22  &    180      &  $-''-$       &  $-''-$        &  $-''-$      &  56300109 \\
   \, PHT22  &    180      &  13 42 32.2   & 40 29 12      & 15$\times$15 &  21301803 \\
   \, PHT25  &    90       &   13 43 16.3  &   40 28 46.5   &     1  &    55800226 \\
   \, PHT25  &    90       &   13 43 16.3  &   40 28 46.5   &     1  &    55800234 \\
   \, PHT25  &   150       &   13 43 16.3  &   40 28 46.6   &     1  &    56200127 \\
   \, PHT25  &   150       &   13 43 16.3  &   40 28 46.5   &     1  &    56200135 \\
   \, PHT25  &   180       &   13 43 16.3  &   40 28 46.7   &     1  &    56300129 \\
   \, PHT25  &   180       &   13 43 16.3  &   40 28 46.5   &     1  &    56300137 \\

NGP(S)          &             &               &                & \\
   \, PHT22  &     90      &  13 49 44.5   & 39 07 37      & 32$\times$4 & 55800215 \\
   \, PHT22  &    150      &  $-''-$       &  $-''-$        &  $-''-$      & 56200114 \\
   \, PHT22  &    180      &  $-''-$       &  $-''-$        &  $-''-$      & 56300113 \\       
       
\hline
\end{tabular}
\label{table:tdt}
\end{table}

\begin{table*}
\caption{%
Observations used for the determination of the zodiacal light emission. The
columns are: (1) name of the field, (2)-(3) position, (4) wavelength, (5) the
ISO identifier number (TDT) of the observation, and (6) time difference
between the listed observation and the observation of the EBL raster maps of
Table~\ref{table:tdt}. A time difference is quoted only when the observations
were not performed within the same day. Observations at wavelengths below
60\,$\mu$m are made with the ISOPHOT-P detector.
}
\begin{tabular}{lcccccc}
\hline  \hline
Field   &    RA       &  DEC       &  $\lambda$ & AOT  & TDT number & $\Delta t$ \\
        &   (J2000)   & (J2000)    &   ($\mu$m) &      &            &  (days)    \\
\hline
EBL22   &  2 26 34.4  & -25 53 53  &    25       & PHT05 & 82101132   &   +2       \\
        &             &            &    90       & PHT25 & 82101126   &   +2       \\
        &             &            &   150       & PHT25 & 81901927   &            \\
        &             &            &   180       & PHT25 & 81901529   &            \\
EBL26\_ZL1$^1$   &  1 20 3.0   &  1 32 30    &  7.3  & PHT05     & 58600239   &    \\
        &             &            &   25        & PHT05 & 58600240   &            \\
        &             &            &   60        & PHT25 & 58600341   &            \\
        &             &            &  170        & PHT25 & 58600444   &            \\
EBL26\_ZL2   &  1 19 30.2  &  1 23 59  &  3.6    & PHT05 & 58600646   &            \\
        &             &            &   7.3       & PHT05 & 58600647   &            \\
        &             &            &   25        & PHT05 & 58600648   &            \\        
        &             &            &   60        & PHT25 & 58600349   &            \\
        &             &            &   90        & PHT25 & 58600342   &            \\
        &             &            &   90        & PHT25 & 58600350   &            \\
        &             &            &  150        & PHT25 & 58600443   &            \\
        &             &            &  150        & PHT25 & 58600451   &            \\
        &             &            &  170        & PHT25 & 58600452   &            \\        
        &             &            &  180        & PHT25 & 58600545   &            \\        
        &             &            &  180        & PHT25 & 58600553   &            \\
NGP\_ZL1  & 13 43 16.3  &  40 28 47  &   7.3     & PHT05 & 55800331   &   -4         \\
        &             &            &   25        & PHT05 & 55800332   &   -4       \\
        &             &            &   60        & PHT25 & 55800233   &   -4       \\
        &             &            &   90        & PHT25 & 55800226   &   -4       \\
        &             &            &   90        & PHT25 & 55800234   &   -4       \\
        &             &            &  150        & PHT25 & 56200127   &           \\
        &             &            &  150        & PHT25 & 56200135   &           \\
        &             &            &  170        & PHT25 & 56200136   &            \\
        &             &            &  180        & PHT25 & 56300129   &    +1       \\
        &             &            &  180        & PHT25 & 56300137   &    +1       \\
NGP\_ZL2$^2$  & 13 43 32.4  & 40 33 35 & 7.3 & PHT05 & 55800123   &    -4       \\
        &             &            &   25        & PHT05 & 55800124   &    -4       \\
        &             &            &  170        & PHT25 & 56200128   &            \\
\hline        
\end{tabular}

$^1$The position is outside the raster maps. \\
$^2$The position is outside maps other than the 180\,$\mu$m map
consisting of 15$\times$15 rasters (TDT number 21301803).

\label{table:ZL_observations}
\end{table*}

Each field was mapped in the PHT22 staring raster map mode (ISOPHOT Handbook,
Laureijs et al. \cite{Handbook}) using filters
C\_90, C\_135, and C\_180. The corresponding reference wavelengths of the
filters are 90\,$\mu$m, 150\,$\mu$m, and 180\,$\mu$m. The 90$\mu$m
observations were made with the C100 detector consisting of 3$\times$3 pixels,
with 43.5$\arcsec \times$43.5$\arcsec$ each. The longer wavelength
observations were made with the C200 detector which has $2\times 2$ detector
pixels, with 89.4$\arcsec \times$89.4$\arcsec$ each. The same raster maps were
used in Juvela et al. (\cite{Juvela00}). Table~\ref{table:fields} lists the
coordinates and the sizes of the maps.
Additionally, we make use of PHT25 absolute photometry measurements (see
ISOPHOT Handbook, Laureijs et al. \cite{Handbook}) made at the same three
wavelengths.
Two positions in NGP, two positions in EBL26, and one position in EBL22 were
observed in this mode.

\begin{table*}
\caption{The positions and sizes of the observed fields. Columns are: (1) name
of the field, (2)-(3) equatorial coordinates of the centre of the field,
(4)-(5) galactic coordinates, (6)-(7) ecliptic coordinates, (8) number of
raster points, (9) area in square degrees, and (10) additional remarks. All
areas were observed at 90, 150, and 180\,$\mu$m. In NGP(N) an additional
square map was observed at 180\,$\mu$m only. Details of the individual
measurements are listed in Appendix, in Table~\ref{table:tdt}. }
\begin{tabular}{llllllllll}
\hline \hline
Field   &  Ra (2000) &  Dec (2000)  &  $l$ & $b$  &   $\lambda$.  &  $\beta$
        & Rasters    &  Area ($\square\degr$) & Notes \\
\hline
EBL22   &  02 26 34.5   &  -25 53 43  & 215.0 & -68.7 &   23.8 & -37.9  &  32$\times$3   & 0.19   
        &  \\
        &  02 32 48.0   &  -25 54 06  & 215.6 & -67.3 &   25.5 & -38.5  &  31$\times$1   & 0.07  
        &  1D scan into a region of higher cirrus \\
EBL26   &  01 18 14.5   &   01 56 40  & 136.5 & -60.2 &   18.8 &  -5.9  &  32$\times$4   & 0.27 &  \\
NGP(N)  &  13 43 53.0   &   40 11 35  &  86.5 &  73.0 &  184.2 &  46.5  &  32$\times$4   & 0.27 
        &  \\
        &  13 42 32.2   &   40 29 12  &  87.9 &  73.0 &  183.6 &  46.6  &  15$\times$15  & 0.56
        &  180\,$\mu$m only\\
NGP(S)  &  13 49 43.7   &   39 07 30  &  81.3 &  72.9 &  186.4 &  46.1  &  32$\times$4   & 0.27   \\
\hline
\end{tabular}
\label{table:fields}
\end{table*}

\subsection{Reduction of EBL field observations} \label{sect:isophot_reduction}

The ISOPHOT data were processed with PIA (PHT Interactive Analysis) program
version 11.3. For details of the analysis steps, see the ISOPHOT Handbook
(Laureijs et al. \cite{Handbook}) and Appendix, section~\ref{sect:techcal}. 
For C100 a method of signal transient correction was
introduced in PIA 11.3. This procedure was used for all C100 measurements.
Nevertheless, some of the internal calibrator (FCS) measurements show residual
drifts. In those cases we applied transient recognition which removes the
initial, unstabilised part of the measurements. The flux density calibration
was made using the internal calibrator measurements (FCS1) performed
immediately before and after each map for actual detector response assessment.
The calibration was applied using the average response of the two FCS
measurements. 

The reduced data contained a few artifacts. These include short time scale
detector drifts at the beginning of some C100 observations, temporary signal
variations caused by cosmic ray glitches, and occasional drifting of some
detector pixels that may also be connected with cosmic ray hits. The time
ordered data were examined by eye. For rasters and detector pixels affected by
clear anomalies (glitches or drifting) the corresponding PIA error estimates
were scaled upwards, typically by a factor of a few.
For each detector pixel the signal values were scaled so that their average
value over a map became equal to the overall average over all detector pixels.
The scaling takes into account the already manually adjusted error estimates.
The flat fielding would actually not be necessary, because FIR fluxes are
compared only with observed HI 21cm lines and, therefore, averaged over areas
that are large compared with the size of the ISOPHOT rasters.

Long term detector response drifts are not taken out by a simple averaging of
the FCS measurements, nor is an initial non-linear drift corrected for by
linear interpolation between the two FCS measurements. Both could introduce an
artificial gradient in the time ordered data and, because of the systematic
scan pattern, also in the maps. The maps were compared with IRAS data in order
to see if there were any gradients uncorrelated with the IRAS 100\,$\mu$m
signal. The only significant difference was found in the C200 observations of
the southern NGP field. The gradient was removed while keeping the average
surface brightness unchanged. The correction has little effect on the
subsequent analysis. Apart from the EBL22 field, all maps contain four
detector scans that run alternatively in opposite directions along the longer
map dimension. When data are correlated with the lower resolution HI
observations, the subsequent scan legs tend to cancel out any long term
drifts.

The raster map observations themselves do not contain any direct measurement
of the dark current. In such cases one usually relies on the orbit dependent
``default'' dark current estimates included in the PIA.  However, absolute
photometry PHT-25 measurements were carried out within a couple of hours
before or after each raster map. The data reduction was carried out also using
the dark current and cold FCS values obtained from those measurements. In the
subsequent analysis, we use maps that are averages of those obtained using
default dark current values and those obtained using PHT-25 dark current
measurements.

When absolute photometry points were inside the mapped area they were compared
with the surface brightness of the raster maps. The maps were re-scaled so
that the final surface brightness corresponds to the average of the original
FCS calibrated maps and the values given by the absolute photometry
measurements. This causes systematic lowering of the surface brightness values
of the original maps. For EBL26, NGP(N), and NGP(S) the change is typically
$\sim$4\%, for both C100 and C200 observations. In the case of EBL22 the
correction is larger, some 20\%, for the C200 detector.

In the region NGP there are separate northern and southern fields that overlap
by a few arc minutes. The maps, each containing 32$\times$4 raster points, were
fitted together using the overlapping area, where the final map is at a level
equal to the average of the northern and the southern maps.  The resulting
change in the surface brightness levels of individual maps was $\sim$5\% or
less. In the north there is yet another 15$\times$15 raster map that was
observed only at 180$\mu$m. Because that measurement includes only very short
FCS measurements, it was scaled to fit the already combined long 180$\mu$m
map. This required scaling of the surface brightness values by a factor of
1.05.

The main maps of the field EBL22 cover an area of low cirrus emission. There
are additional one-directional scans that extend to a region of higher surface
brightness in the west. In the absence of scans in the opposite direction, it
is not possible to directly determine the presence of detector response
drifts. However, these observations were reduced using the average of the
responsivities given by the two FCS measurements and the error bars reflect
also the difference in the responsivity before and after the measurement.
Using the overlapping area, the $32\times 1$ raster strips were scaled to the
same level with the $32\times 3$ raster maps. The scalings applied were 0.97,
1.02, and 0.84 at 90\,$\mu$m, 150\,$\mu$m, and 180\,$\mu$m, respectively.

The final FIR errorbars show the uncertainty for the weighted means over the
Effelsberg beam. The noise of each HI spectrum was estimated separately using
the velocity channels outside the line. The uncertainty of the line area was
calculated assuming the same, uncorrelated noise for the integrated velocity
interval. This might underestimate the total uncertainty, because it ignores
the uncertainties in the stray radiation subtraction that do not affect the
signal in the line wings. However, for a small field the stray radiation
causes a constant systematic error rather than statistical uncertainty and
does not affect the weighting of the observations when the linear fit is made.

\begin{figure*} 
\resizebox{18cm}{!}{\includegraphics{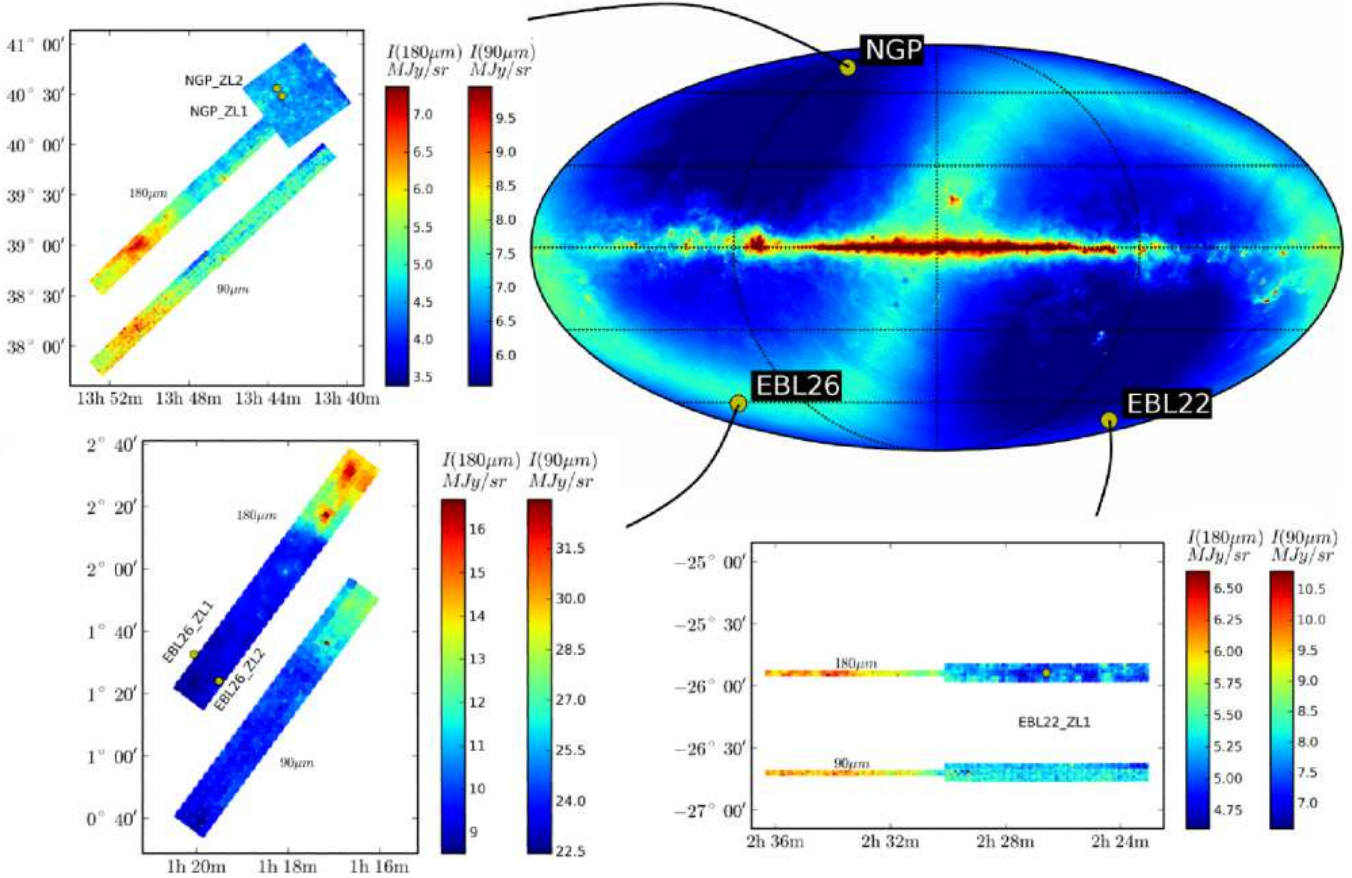}}
\caption{
The ISOPHOT EBL fields. The three frames show the 180\,$\mu$m and 90\,$\mu$m
ISOPHOT maps. The coordinates correspond to the 180\,$\mu$m maps. The 
90\,$\mu$m maps cover the same area but, in the figure, the 90\,$\mu$m maps
have been plotted south of the 180\,$\mu$m maps. The small yellow circles
indicate the positions observed with the ISOPHOT P-detector for the
determination of the zodiacal light levels. To indicate the locations of the
fields with respect to the galactic and ecliptica planes, the positions are
shown on an all-sky map that is combined from DIRBE observations between the
wavelengths of 12\,$\mu$m and 240\,$\mu$m. 
} 
\label{fig:allsky}
\end{figure*}

For selected positions there exist mid-infrared observations made with the
ISOPHOT P-detectors as well as further absolute photometry measurements with
the C100 and C200 cameras (see Appendix, Table~\ref{table:ZL_observations}).
These observations were performed for the purpose of estimating the
zodiacal light. The data reduction of P-detector data is similar to that of
the C100 and C200 cameras, except that also signal linearisation is included.

\subsection{HI measurements} \label{sect:HImeas}

\begin{figure} 
\resizebox{8cm}{!}{\includegraphics{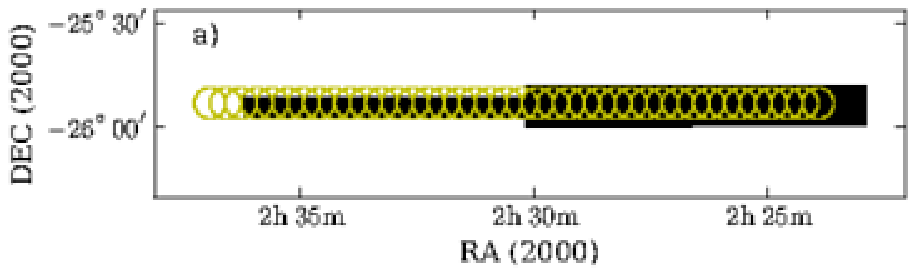}}
\resizebox{8cm}{!}{\includegraphics{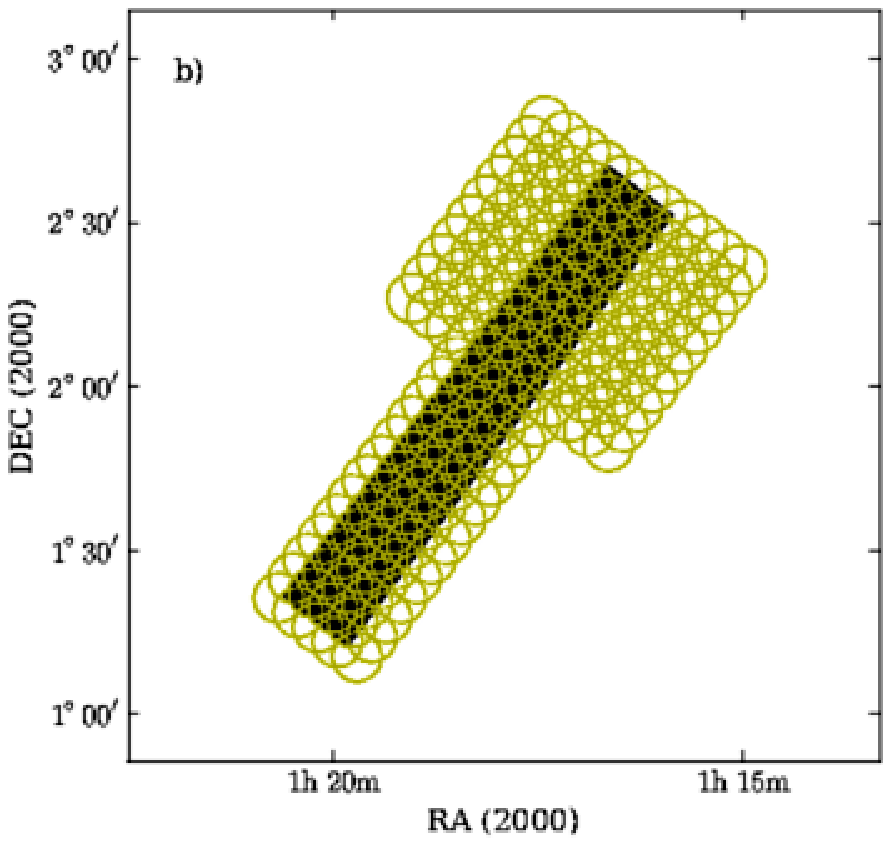}}
\resizebox{8cm}{!}{\includegraphics{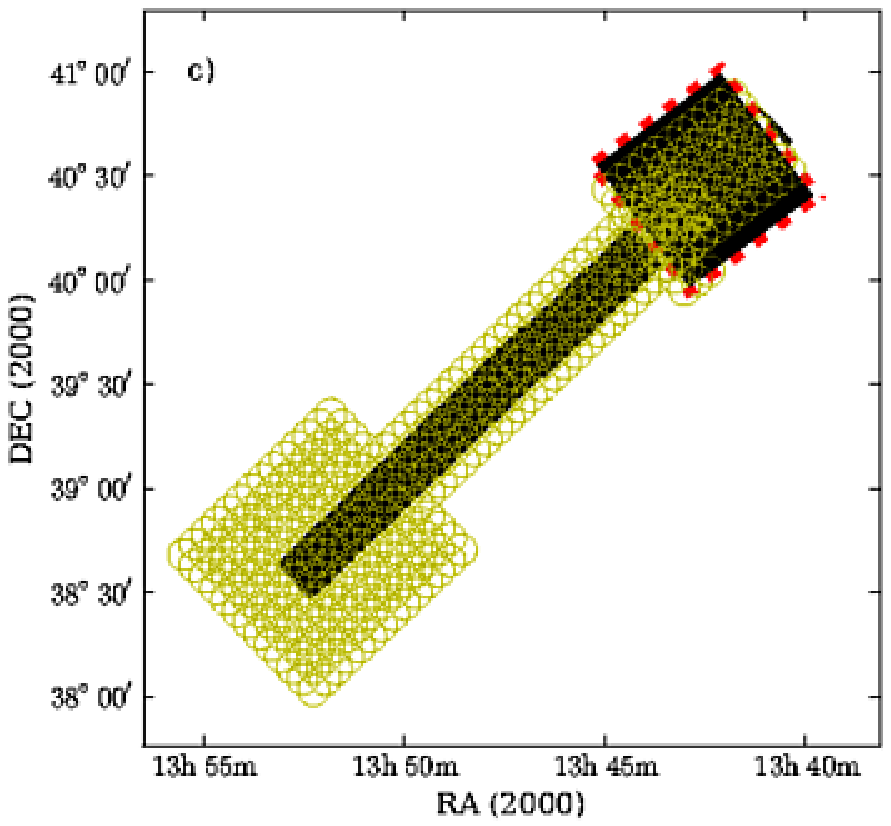}}
\caption{
The figures shows as black rectangles the areas mapped with ISOPHOT (90, 150,
and 180\,$\mu$m) and as circles the pointings used in the Effelsberg HI
observations. The diameter of the circles, 9$\arcmin$, is equal to the FWHM of
the Effelsberg beam.  The frames $a$, $b$, and $c$ correspond to regions
EBL22, EBL26, and NGP. In the case of NGP, the dashed red line indicates the
area that was mapped at 180\,$\mu$m only.
} 
\label{fig:HI_positions}
\end{figure}

The observations of the hydrogen 21\,cm line were made with the Effelsberg
radio telescope in May 2002. The observed positions, $~580$ in number, are
indicated in Fig.~\ref{fig:HI_positions}. The integration times were 30\,s in
EBL22 and 62\,s in EBL26. In the field NGP the observations were done with
62\,s integrations except for the northern part where the integration time was
94\,s. The average noise estimated from the velocity intervals outside the HI
line is 0.15\,K per channel of 1\,km/s. This corresponds to a typical
uncertainty of 1.7\,K\,km\,s$^{-1}$ in the integrated line area. 

For calibration purposes and for precise subtraction of the stray radiation,
regular observations of the standard region S7 were made. The stray radiation
subtraction is crucial because it affects the zero point of the estimated HI
column densities. The observed fields, NGP in particular, have some of the
lowest line-of-sight column densities over the whole sky. Under these
conditions the stray radiation received by the telescope side lobes becomes a
significant fraction of the total signal. The stray radiation was removed with
a program developed by P. Kalberla (Kalberla et al. \cite{Kalberla2005}; see
Sect.~\ref{sect:obs}).

\begin{figure} 
\resizebox{\hsize}{!}{\includegraphics{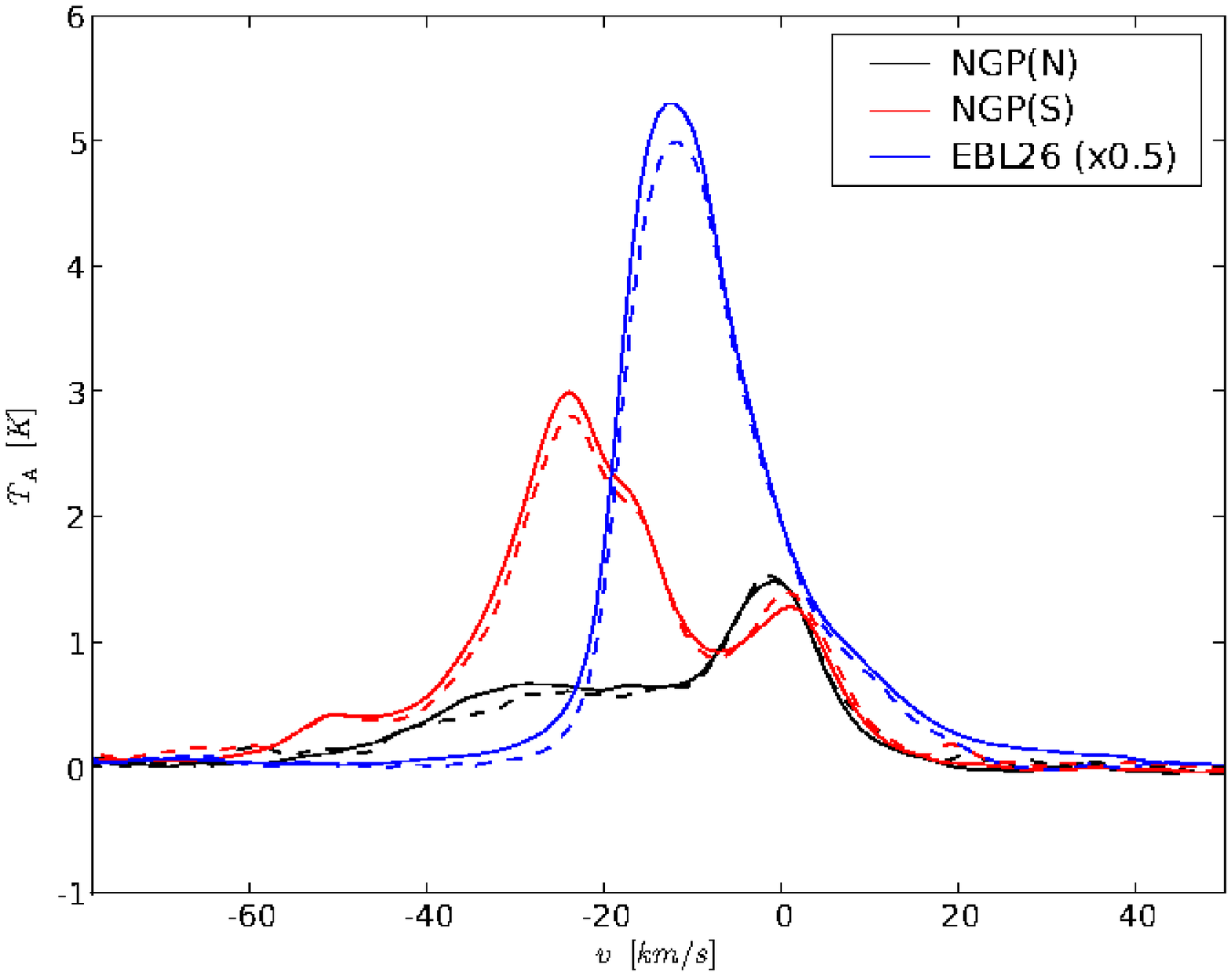}}
\caption{
Comparison of HI spectra from the Leiden/Dwingeloo survey (Kalberla et al.
\cite{Kalberla2005}; dashed lines) and our Effelsberg data convolved with a
beam of 36$\arcmin$ (solid lines). The spectra correspond to positions at the
southern and northern end of the NGP map (13$^{\rm h}$\,42$^{\rm m}$\,0$^{\rm
s}$ +40$\degr$\,30$\arcmin$\,0$\arcsec$ and 13$^{\rm h}$\,52$^{\rm
m}$\,0$^{\rm s}$ +38$\degr$\,40$\arcmin$\,0$\arcsec$) and one position in the
field EBL26 (1$^{\rm h}$\,17$^{\rm m}$\,0$^{\rm s}$
+2$\degr$\,20$\arcmin$\,0$\arcsec$). The EBL26 spectra have been scaled by a
factor 0.5.
} 
\label{fig:HI_comparison}
\end{figure}

In Fig.~\ref{fig:HI_comparison} we compare our data with spectra from the 
Leiden/Dwingeloo survey (Hartmann \& Burton \cite{Hartmann1997}; Kalberla et
al. \cite{Kalberla2005}). For this comparison, in order to match the
resolution of the Leiden/Dwingeloo survey, the Effelsberg data were convolved
with a gaussian with FWHM equal to 36$\arcmin$. The HI profiles agree very
well. Part of the differences may be caused by the fact that our HI maps do
not cover the whole area of the 36$\arcmin$ beams. Nevertheless, the figure
shows that the observations and the stray radiation subtraction (see
Sect.~{sect:obs}) are consistent with the Kalberla (\cite{Kalberla2005})
results.

\section{Calibration accuracy} \label{sect:calibration}

The error estimates listed in Table~\ref{table:fit} are based on the
statistical uncertainties in the fits between FIR and HI data. 
The scatter of data points around the fitted lines is usually larger than
their estimated uncertainty. This could be a sign of underestimated
measurement uncertainties but is more likely caused by true scatter in the
relation. If the formal uncertainties of the line parameters were estimated
based on the error estimates of the individual points, the uncertainties could
be severely underestimated. Therefore, instead of relying only on the
measurement uncertainties, the uncertainty of the fit parameters was estimated
separately with the bootstrap method so that they reflect the true scatter of
observed points. The error estimates corresponding to a 67\% confidence
interval are given in Table~\ref{table:fit}. 

These uncertainties do not include estimates for the systematic errors
introduced by the independent calibration of each map or the absolute accuracy
of the overall ISOPHOT calibration. There are both multiplicative and additive
sources of uncertainty. The former include, for example, uncertainties in the
internal calibration source (FCS) measurements (e.g., detector drifts) that
alter the estimated detector response. The uncertainties that affect the zero
point of the intensity scale are more critical, because the CIRB is small
compared with the observed signal and can be recovered only as the residual
after the subtraction of the ZL. 

Table~\ref{table:uncertainties} lists an assessment of uncertainty that, using
data in Table~\ref{table:fit}, have been converted into uncertainty of the FIR
flux at zero HI column density. The quoted values are half of the difference
of two values obtained in two independent ways. Thereby the quoted values are
also $\sim$1-$\sigma$ estimates for the uncertainty of the average of the two
values.

In Table~\ref{table:uncertainties} column 4 has been obtained by comparing the
fine calibration source measurements performed before and after each map. The
numbers indicate the statistical uncertainty of the detector response measurements.
The FCS measurements are generally very consistent, particularly in the case
of the C200 detector. On the other hand, the effect of the drift affecting the
first FCS measurement of the one-dimensional strip map of EBL22 is 
clearly visible at 90$\mu$m.

The dark signal subtraction is the most important correction affecting the
zero point of the FIR intensity. Close to each of the raster map observations,
we have one or two absolute photometry observations which include dark signal
measurements of their own. In PIA, the default dark current calibration is
based on a larger set ($\sim$70) of dark current measurements for which the
orbit trend has been determined. Therefore, the PIA default dark current
calibration is less affected by the noise of individual measurements but may
not take into account short time scale variations in the detector dark current
on a specific orbit. The maps were reduced using the default dark current
values and the actually measured dark current values. In
Table~\ref{table:uncertainties} column 5 shows the associated uncertainty in
the FIR signal at zero HI column density.
The observed uncertainty in the dark current values is comparable with the
variation observed in the systematic analysis of a large sample of ISOPHOT
observations (del Burgo et al. \cite{delBurgo02}; see also
Fig.~\ref{fi:c100pix5darksignal}).

When absolute photometry measurements existed within mapped areas, those were
used to re-scale the surface brightness values of the maps (see
Sect.~\ref{sect:isophot_reduction}). The difference in the absolute photometry and
mapping measurements is used to derive the values in column 6 of
Table~\ref{table:uncertainties}. The final column reflects the difference in
the surface brightness in areas where two independently calibrated maps
overlap. The numbers in columns 6 and 7 include, of course, dark current and
FCS uncertainties as one of their components. 
For the C100 observations at 90\,$\mu$m the uncertainty is close to
1\,MJy\,sr$^{-1}$, i.e., comparable with the expected EBL signal. On the other
hand, for the C200 detector the uncertainty of an individual map is
$\sim$0.3\,MJy\,sr$^{-1}$. Most of this is caused by the uncertainty in the
dark current values.

\begin{table*}
\caption{Assessment of the calibration uncertainty for the ISOPHOT maps. Columns
are (1) name of the field, (2) wavelength, (3) average surface brightness of
the map, (4) difference between calibration measurements performed before and
after each map, (5) difference between actual dark current
measurements and default dark current values, (6) difference between the 
independently calibrated absolute
photometry measurements and raster maps, and (7) difference between partially
overlapping maps. These uncertainties have been converted to correspond to the uncertainties at zero
hydrogen column density using the fit parameters listed in Table~\ref{table:fit}.
}
\begin{tabular}{lrccccc}
\hline \hline
   Field    &  $\lambda$    
            &    $<S>$                  
            &  $\Delta$(FCS)   
            &  $\Delta$(DC)
            &  $\Delta$(Abs.)              
            &  $\Delta$(Join) \\
            & ($\mu$m)                    
            &   MJy\,sr$^{-1}$     
            &   MJy\,sr$^{-1}$     
            &   MJy\,sr$^{-1}$    
            &   MJy\,sr$^{-1}$       
            &   MJy\,sr$^{-1}$ \\
\hline
        EBL22 &   90 &   9.0 &   0.12 &   0.43 &   -1.15 &   -        \\
        EBL22 &  150 &   5.7 &  -0.06 &   1.44 &   -0.64 &   -        \\
        EBL22 &  180 &   4.5 &  -0.38 &   0.48 &   -0.28 &   -        \\
    EBL22$^1$ &   90 &   6.5 &  -1.09 &   0.22 &   -     &  -0.17     \\
    EBL22$^1$ &  150 &   3.6 &   0.02 &   0.30 &   -     &   0.07     \\
    EBL22$^1$ &  180 &   4.5 &   0.04 &   1.00 &   -     &  -0.50     \\
        EBL26 &   90 &  20.6 &  -0.09 &  -0.80 &   -1.66 &       -    \\
        EBL26 &  150 &   4.3 &  -0.05 &  -0.28 &   -0.09 &       -    \\
        EBL26 &  180 &   3.7 &   0.06 &  -0.20 &    0.11 &       -    \\
       NGP(N) &   90 &   7.8 &   0.49 &  -0.19 &   -0.50 &   0.58     \\
       NGP(N) &  150 &   5.1 &   0.06 &  -0.30 &   -0.35 &   0.19     \\
       NGP(N) &  180 &   4.8 &   0.22 &  -0.22 &   -0.26 &   0.39     \\
       NGP(S) &   90 &   6.9 &  -0.32 &   0.26 &   -     &  -0.49     \\
       NGP(S) &  150 &   4.7 &   0.04 &  -0.30 &   -     &  -0.18     \\
       NGP(S) &  180 &   4.7 &   0.05 &  -0.31 &   -     &  -0.31     \\

\hline
{$^1$The narrow strip map.}
\end{tabular}
\label{table:uncertainties}
\end{table*}

\subsection{Straylight radiation}

Straylight may be another instrumental artefact affecting the zero level
of the FIR surface brightness. By design and operation ISO's viewing direction
stayed by several tens of degrees away from the brightest FIR
emitters in the sky, the Sun, the Earth and the Moon (Kessler et al.
\cite{Kessler2003}). A dedicated straylight program was executed verifying by
deep ``differential'' integrations that the uniform straylight level due
to these sources was below ISOPHOT's detection limit, even under the most
unfavourable pointing conditions close to the visibility constraints (Lemke et
al. \cite{Lemke2001}).

Specular straylight by the second brightest class of objects, the giant 
planets Jupiter and Saturn, was observed when pointing to within 15$\arcmin$
to 1$\degr$ of the planet, expressing itself as finger-like stripes or faint
ghost rings (Kessler et al. \cite{Kessler2003}, Lemke et al.
\cite{Lemke2001}). The NGP and EBL 22 fields are far away from the ecliptic
and can thus not suffer from this type of straylight. For EBL 26 we checked
the positions of the planets Mars, Jupiter, Saturn, Uranus and Neptune at the
time of the observations, 1997-06-26 and 1997-07-11, respectively. Mars,
Jupiter, Uranus, Neptune were all far off. Saturn was at a distance of 3.25
degrees, which is still more than a factor of 3 off of any known
straylight-critical distance.

\begin{figure} 
\resizebox{8cm}{!}{\includegraphics{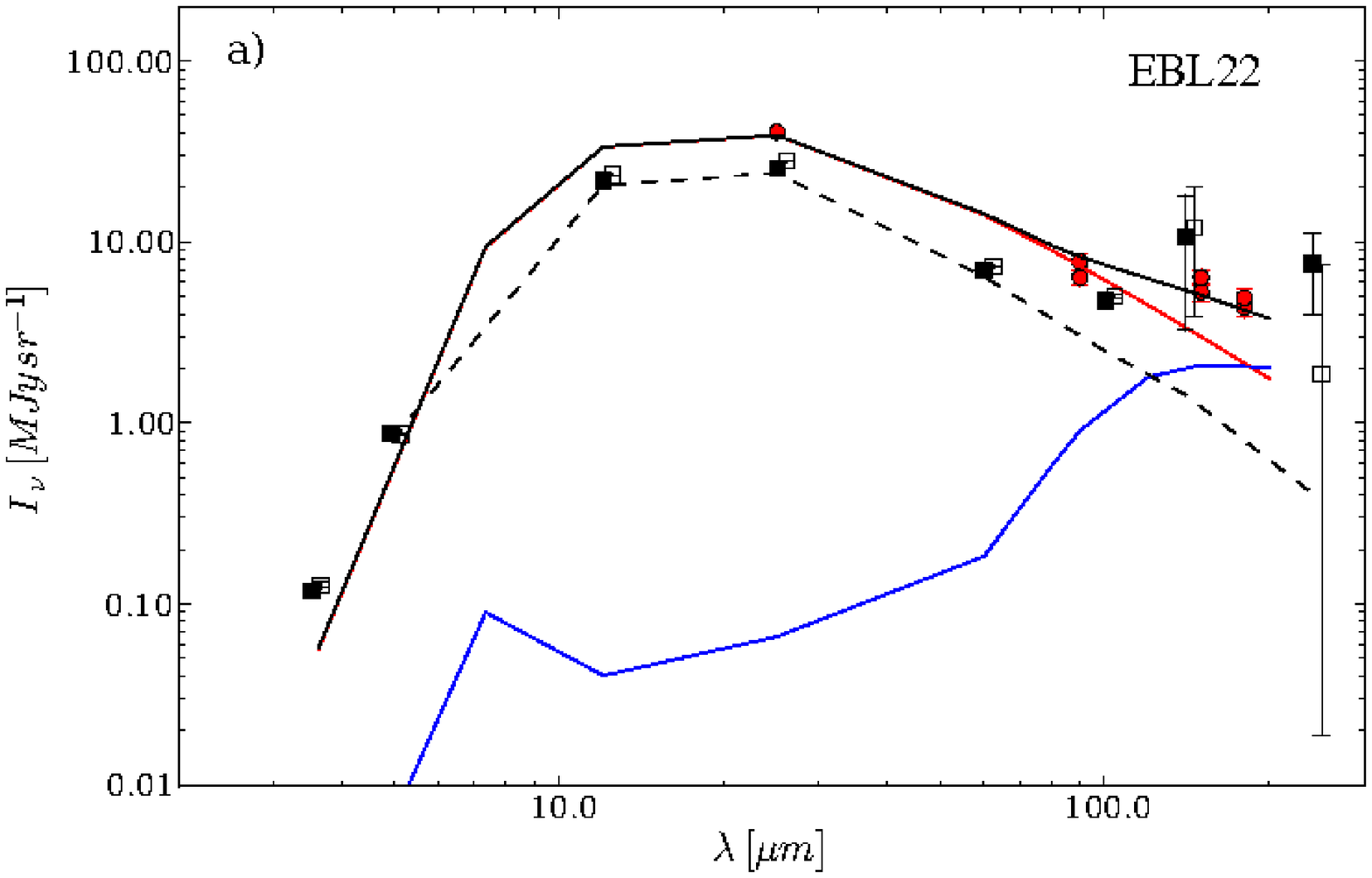}}
\resizebox{8cm}{!}{\includegraphics{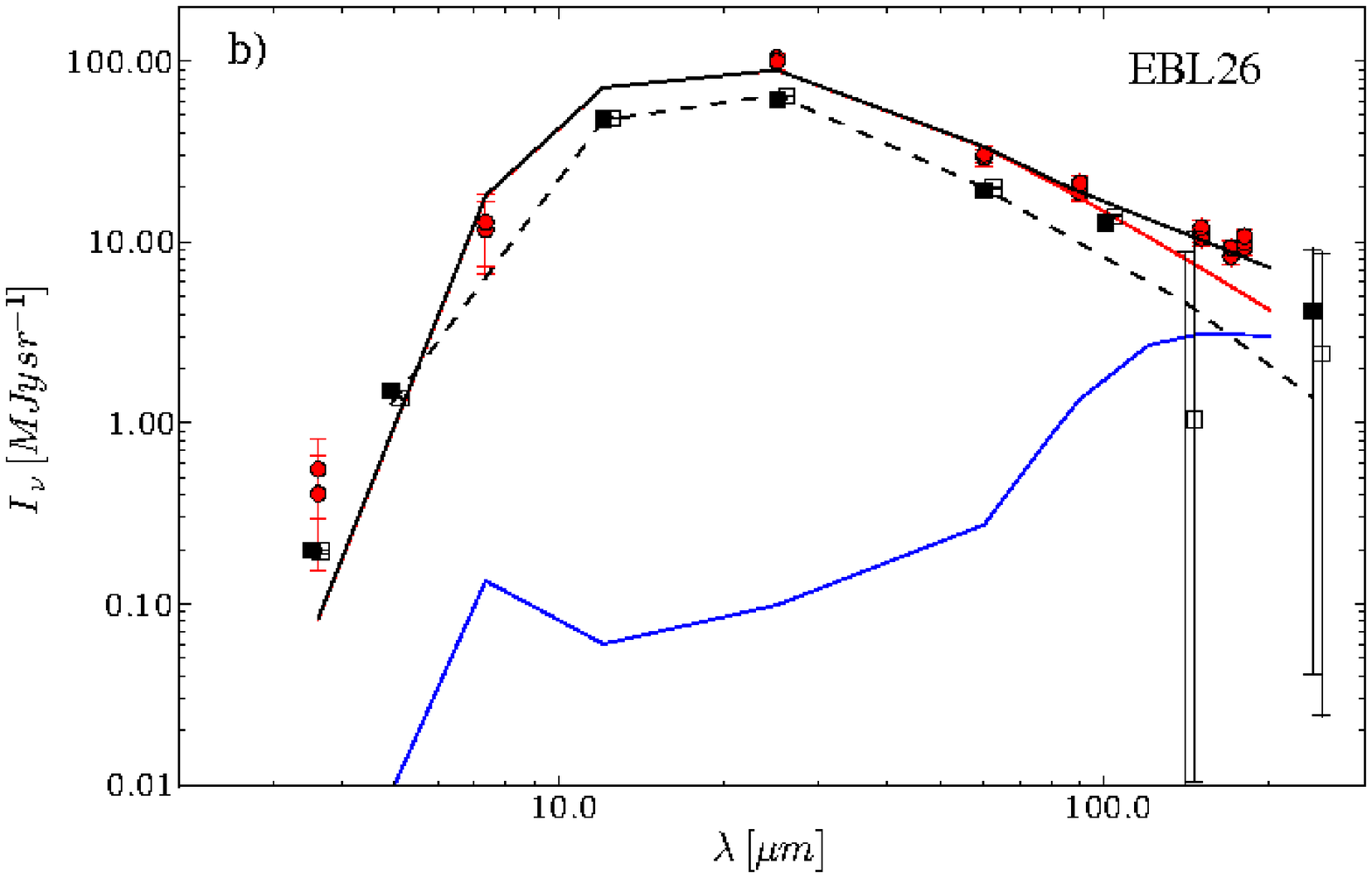}}
\resizebox{8cm}{!}{\includegraphics{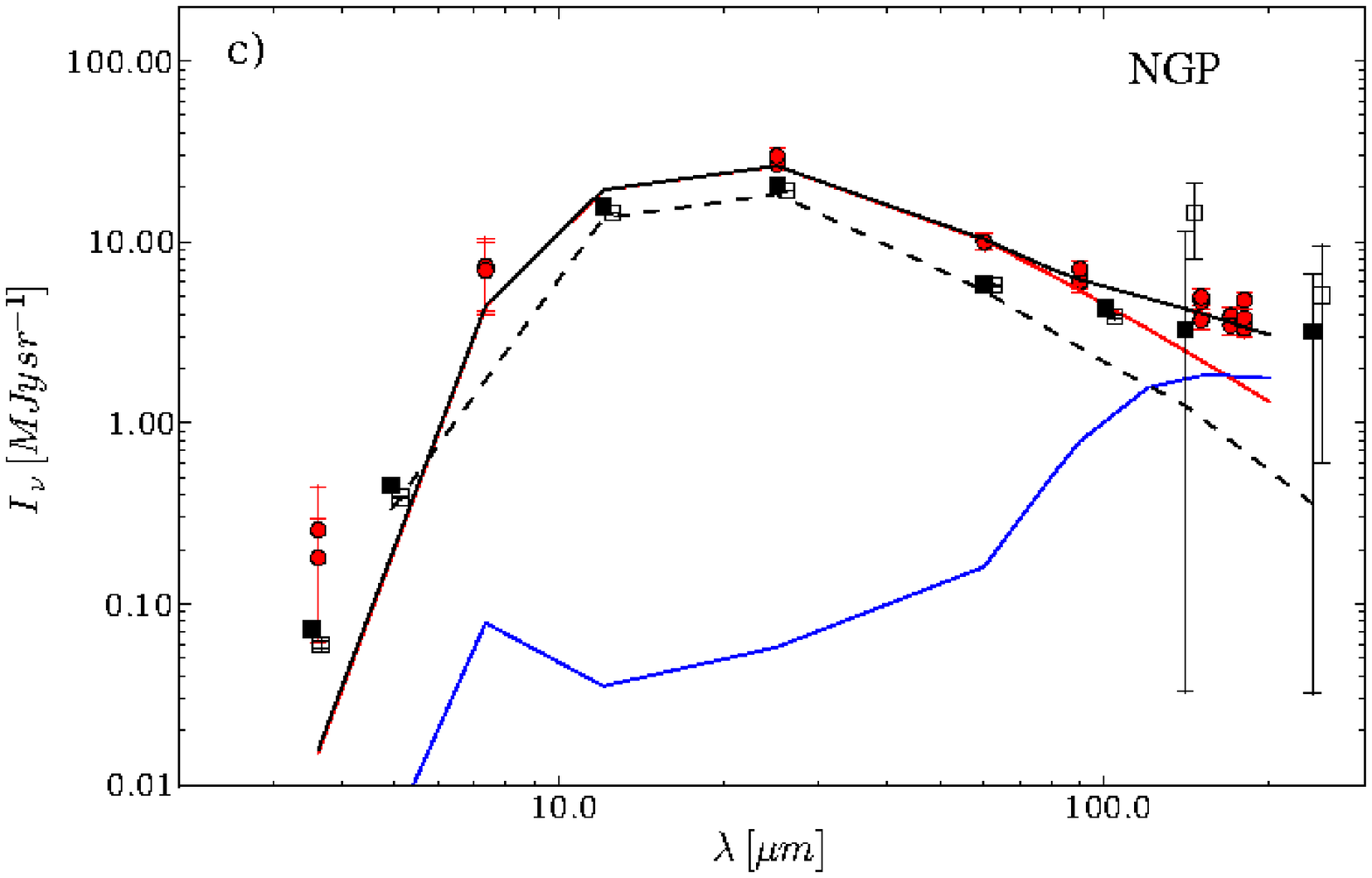}}
\caption{
Fits used to estimate the ZL levels in the three fields EBL22, EBL26, and NGP
(frames a, b, and c, respectively). The red circles are ISOPHOT observations.
The lower lines are the cirrus (blue solid line) and the ZL (red solid line)
templates, the uppermost solid green line is their sum.  The figures also show DIRBE
values for the closest DIRBE pixel, read from the DIRBE weekly maps. The solid
squares correspond to observations with the same solar elongation as in the
case of ISOPHOT observations, the open squares to the other measurement with
identical absolute value of the solar aspect angle but opposite solar
elongation. For clarity, the latter have been shifted slightly in wavelength.
The dashed line shows the predictions of the Kelsall et al. (\cite{Kelsall98})
ZL model.
}
\label{fig:zlfit}
\end{figure}

\section{Determination of the ZL levels} \label{sect:zlfit}

The ZL level was estimated by fitting ZL and cirrus templates to ISOPHOT
observations in the wavelength range from 7.3\,$\mu$m to 200\,$\mu$m (see
Sect.~\ref{sect:ZL}).  Figure~\ref{fig:zlfit} shows the results of these fits.
In the field EBL22 we had observations of one position and in the fields EBL26
and NGP of two positions (see Table~\ref{table:ZL_observations}). For the
latter two fields, the figures show the fit to data combined from the two
positions.

Table~\ref{table:ZL_observations} lists the time difference between the listed
observations and the observations of the raster maps. In the case of NGP these
are relative to the 150\,$\mu$m observations. The 90\,$\mu$m maps were
observed four days before and the 180\,$\mu$m one day after the 150\,$\mu$m
maps. According to the Kelsall et al. (\cite{Kelsall98}) ZL model the four day
difference causes only $\sim$1.5\% change in the expected ZL. The combined
NGP map is almost 1.5 degrees long. In the Kelsall model the difference in the
centre positions of the southern and northern parts corresponds to about 1\%
difference in the ZL. Therefore, we use only one zodiacal estimate value for
both NGP(N) and NGP(S) and for all observations made during the five day
interval.

In the fields EBL26 and NGP, MIR observations exist for two separate positions
(see Fig.~\ref{fig:allsky}). In both fields, the measurements at these two
positions are close to each other, both in time and position. Therefore, their
ZL values should be identical and also the cirrus levels should be very
similar. Comparison of the fits performed using these independent sets of
measurements gives the first indication of the statistical uncertainty of the
ZL values. In both fields, the ZL values obtained for the two positions agree
within 10\%. 

The observations are fitted as a sum of ZL and cirrus components. The ZL
template is a black body curve at the temperature obtained from Leinert et al.
(\cite{Leinert2002}). The cirrus template is based on the model by Li \&
Draine (\cite{Li2001}). Using the ISOPHOT filter profiles we calculate for
both radiation components, ZL and cirrus, and for each filter the in-band
power values that can be directly compared with the observed values. In the
fit we have only two free parameters, the intensity of the ZL component and
the intensity of the cirrus component. The ZL estimates should be based mainly
on data between 10\,$\mu$m and 60\,$\mu$m where the ZL is clearly the dominant
component. Therefore, in the fit, the weight of the data points in this
wavelength range is increased by a factor of two. 

The level of the cirrus component is determined mostly by the longer
wavelength data. In reality, the component corresponds to the sum of the
cirrus and CIRB signals. As long as the component is small in the MIR, the ZL
estimates are almost independent of the exact shape of this template. We
confirmed this by replacing the Li \& Draine (\cite{Li2001}) cirrus template
by a pure CIRB template, using the model curve from Dole et al.
(\cite{Dole2006}; Fig. 13). The resulting change in the ZL estimates was less
than one per cent.

The actual statistical errors of the ZL values are estimated using the
standard deviation of the relative errors when observations are compared with
the fitted ZL curve. The last column of Table~\ref{table:ZL_estimates} lists
the corresponding error of the mean, calculated using data points between
7.3\,$\mu$m and 90\,$\mu$m.  In the case of fields NGP and EBL26, the error
estimates are calculated from the fits where we have combined the data from
the two measured positions within each field. In all three fields, the
obtained relative uncertainties are $\sim$10\%. In the fields EBL26 and
NGP the uncertainties are also consistent with the difference of the ZL
values obtained for the two individual positions.
The ZL fits are shown in Fig.~\ref{fig:zlfit}.

In this paper we have used original ISOPHOT observations without applying
colour corrections. Therefore, in the fitting procedure also the ZL and cirrus
templates were converted to corresponding values using the ISOPHOT filter
profiles. However, for Fig.~\ref{fig:zlfit} we have performed colour
corrections. The templates are plotted by connecting the values at the nominal
wavelengths by a straight line. The template spectra used in the ZL fitting
are colour corrected using their respective spectral shapes. In the figure,
the colour correction of the observed surface brightness values is done
assuming the blackbody ZL spectrum below 90\,$\mu$m, and a modified black body
cirrus spectrum, $B_{\nu}(T=18\,{\rm K}) \nu^2$, at 90\,$\mu$m and longer
wavelengths.

The plots include DIRBE values from the DIRBE weekly maps. These correspond to
the DIRBE pixel closest to the centre of the corresponding ISOPHOT map. 
Linear interpolation was performed between the weeks in order to accurately
match the solar elongation of the ISOPHOT observations. In addition to the
DIRBE value that corresponds directly to the ISOPHOT observations (solid
squares) we plot the DIRBE value for the same solar aspect angle and opposite
sign of the solar elongation. Assuming that the zodiacal dust cloud is
symmetric along the ecliptic, the two values should be identical.  The
predictions of the ZL model of Kelsall et al. (\cite{Kelsall98}) are also
plotted. The DIRBE values are colour corrected. As in the case of ISOPHOT
data, colour correction of the observations assumes a blackbody ZL spectrum at
and below 60\,$\mu$m, and a modified black body cirrus spectrum,
$B_{\nu}(T=18\,{\rm K}) \nu^2$, at the longer wavelengths, 100, 140, and
240\,$\mu$m.

There is a clear difference in the ISOPHOT and DIRBE surface brightness
scales. The DIRBE values are consistently lower by some 20--30\%, in the MIR
range. In the FIR bands the extended cirrus structures combined with the much
larger pixel size and noise in the DIRBE pixels precludes direct comparison.
The determination of the CIRB values is not directly affected by a possible
calibration difference between DIRBE and ISOPHOT because, in this paper, we
use exclusively ISOPHOT measurements. 

Systematic uncertainties affecting all ISOPHOT bands have only little impact
on the derived CIRB values. The relative calibration accuracy between the FIR
cameras and the ISOPHOT-P photometer is more important, because the zodiacal
light estimates are based on the latter. When the absolute level of the
zodiacal light was estimated we calculated the scatter between the SED model
and the observations at different wavelengths (see
Table~\ref{table:ZL_estimates}). The scatter was typically $\sim$10--20\%. The
importance of this error source depends, of course, on the absolute level of
the ZL emission. The field EBL26 is located near the ecliptic plane and at
90\,$\mu$m the observed signal and the ZL are both of the order of
20\,MJy\,sr$^{-1}$. Therefore, a relative uncertainty of 10\% would already
correspond to about twice the expected level of the CIRB. For EBL22 and
especially for NGP the zodiacal light level is much lower so that more
meaningful limits can be derived for the CIRB also at 90\,$\mu$m.

The quoted ZL error estimates reflect the uncertainty in the determined ZL
level in the mid-infrared. If there were a systematic difference in the
calibration of the mid- and FIR-bands, the ZL estimates could be wrong by the
corresponding amount. Generally the relative calibration accuracy is
considered to be within 15\%. This uncertainty would not necessarily be
reflected in the quality of the ZL spectrum fits, because a systematic
calibration error could have been partly compensated by a change in the
intensity of the cirrus component.

The ZL spectrum was assumed to be a pure black body with the temperature given
by Leinert et al. (\cite{Leinert2002}). As far as the mid-infrared points are
concerned, a wrong temperature would, at some level, be reflected also in our
error estimate. However, if the ZL spectrum deviated from the assumed shape
only in the FIR this could again be masked by a change in the fitted cirrus
component without a corresponding increase in the rms value. Therefore, we
must explicitly assume that the same ZL temperature is applicable both
at mid-infrared and far-infrared wavelengths. However, because a 5\,K change in
the ZL temperature corresponds to only $\sim$2\% relative change in the ratio
of 150\,$\mu$m and 25\,$\mu$m intensities, this source of uncertainty is
unimportant compared with the uncertainty in the relative calibrations of the
different detectors.

\section{Comparison with DIRBE EBL estimates} \label{sect:dirbe_ebl}

This present study represents the first determination of the absolute level of
the FIR EBL that is independent of measurements of the COBE DIRBE
instrument. In Table~\ref{table:comparison} we list FIR EBL estimates given in
seven publications based on the DIRBE measurements. Included are also our
2-$\sigma$ upper limit at 90$\mu$m and the EBL estimate for the range
150-180$\mu$m.

\begin{table}
\caption{Comparison of existing CIRB estimates in the FIR range. The
error estimates quoted by the authors are shown in parenthesis. In our case, we
include only the statistical uncertainty. }
\begin{tabular}{rcll} 
\hline \hline
$\lambda$ &   $I_{\nu}$        &  Reference   & Instrument \\
($\mu$m)  &   (MJy\,sr$^{-1}$) &              & \\
\hline
90        &   $<$2.3        &  this paper      & ISO/ISOPHOT \\
150/180   &   1.08 (0.32)$^2$ &  this paper      & ISO/ISOPHOT \\

100       &   $<1.1^1$      &  Hauser et al. 1998  & COBE/DIRBE \\
          &   0.73 (0.20)$^1$  &    &  \\
100       &   0.37 (0.10)  &  Dwek et al. 1998  & COBE/DIRBE \\
100       &   0.78 (0.20)  &  Lagache et al. 2000  & COBE/DIRBE \\
100       &   0.83 (0.27)  &  Finkbeiner et al. 2000  & COBE/DIRBE \\

140       &   1.49 (0.33)  &  Schlegel et al. 1998  & COBE/DIRBE \\
140       &   1.17 (0.33)  &  Hauser et al. 1998  & COBE/DIRBE \\
140       &   0.70 (0.28)  &  Hauser et al. 1998  & COBE/FIRAS \\
140       &   0.70 (0.28)  &  Lagache et al. 1999  & COBE/DIRBE \\
140       &   1.12 (0.56)  &  Lagache et al. 2000  & COBE/DIRBE \\
140       &   1.17 (0.37)  &  Odegard et al. 2007 & COBE/DIRBE \\

240       &   1.36 (0.16)  &  Schlegel et al. 1998  & COBE/DIRBE \\
240       &   1.12 (0.24)  &  Hauser et al. 1998  & COBE/DIRBE \\
240       &   1.04 (0.16)  &  Hauser et al. 1998  & COBE/FIRAS \\
240       &   0.88 (0.16)  &  Lagache et al. 1999  & COBE/DIRBE \\
240       &   0.88 (0.56)  &  Lagache et al. 2000  & COBE/DIRBE \\
240       &   1.04 (0.24)  &  Odegard et al. 2007 & COBE/DIRBE \\

\hline
\end{tabular}
$^1$ Hauser et al. did not claim detection at 100\,$\mu$m, 
because the CIRB signal failed test for isotropy. \\
$^2$Only the statistical error is quoted.
\label{table:comparison}
\end{table}

\end{document}